\documentclass[twocolumn]{aastex63}\tighten

\usepackage{lineno}
\usepackage{amsmath}
\received{\textit{November 30, 2020}}
\revised{\textit{February 18, 2021}}
\accepted{\textit{February 23, 2021}}

\submitjournal{ApJ}

\shorttitle{Magnetic field in the S235 complex}
\shortauthors{Devaraj et al.}

\def\Msun{\hbox{M$_{\odot}$}}

\def\13CO{$^{13}$CO}

\begin{document}

\title{\large Magnetic fields and Star Formation around H\,{\sc ii} regions: The S235 complex}

\correspondingauthor{Devaraj Rangaswamy}
\email{dev2future@yahoo.com}

\author[0000-0001-9217-3168]{Devaraj R.}
\affiliation{Dublin Institute for Advanced Studies, 31 Fitzwilliam Place, Dublin D02XF86, Ireland}

\author[0000-0002-9947-4956]{D.P. Clemens}
\affiliation{Institute for Astrophysical Research, Boston University, 725 Commonwealth Avenue, Boston, MA 02215, USA}

\author[0000-0001-6725-0483]{L.K. Dewangan}
\affiliation{Physical Research Laboratory, Navrangpura, Ahmedabad, Gujarat 380009, India}

\author[0000-0002-3922-6168]{A. Luna}
\affiliation{Instituto Nacional de Astrof\'isica, \'Optica y Electr\'onica, Tonantzintla, Puebla 72840, M\'exico}

\author[0000-0002-2110-1068]{T.P. Ray}
\affiliation{Dublin Institute for Advanced Studies, 31 Fitzwilliam Place, Dublin D02XF86, Ireland}

\author[0000-0002-5449-6131]{J. Mackey}
\affiliation{Dublin Institute for Advanced Studies, 31 Fitzwilliam Place, Dublin D02XF86, Ireland}
\affiliation{Centre for Astroparticle Physics and Astrophysics (CAPPA), DIAS Dunsink Observatory, Dunsink Lane, Dublin 15, Ireland}
%\author{et al.}

\begin{abstract} %currently 278 words

Magnetic fields are ubiquitous and essential in star formation. In particular, their role in regulating formation of stars across diverse environments like H\,{\sc ii} regions needs to be well understood. In this study, we present magnetic field properties towards the S235 complex using near-infrared (NIR) $H$-band polarimetric observations, obtained with the Mimir and POLICAN instruments. We selected 375 background stars in the field through combination of \textit{Gaia} distances and extinctions from NIR colors. The plane-of-sky (POS) magnetic field orientations inferred from starlight polarization angles reveal a curved morphology tracing the spherical shell of the H\,{\sc ii} region. The large-scale magnetic field traced by {\it Planck} is parallel to the Galactic plane. We identified 11 dense clumps using $1.1\,\mathrm{mm}$ dust emission, with masses between $33-525\,\rm M_\odot$. The clump averaged POS magnetic field strengths were estimated to be between $36-121\,\mathrm{\mu G}$, with a mean of ${\sim}65\,\mathrm{\mu G}$. The mass-to-flux ratios for the clumps are found to be sub-critical with turbulent Alfv\'{e}n Mach numbers less than 1, indicating a strongly magnetized region. The clumps show scaling of magnetic field strength vs density with a power-law index of $0.52\pm0.07$, similar to ambipolar diffusion models. Our results indicate the S235 complex is a region where stellar feedback triggers new stars and the magnetic fields regulate the rate of new star formation. 

\end{abstract}


\section{Introduction}

The advent of space based mid-infrared observations has revealed parsec-scale ring-like structures around ionized regions \citep{church06,deharveng10}. These are formed by massive stars through their energetic winds, UV ionizing radiation, and expanding H\,{\sc ii} regions \citep{zinnecker07}. The resulting turbulent interaction with gas and dust sweeps up the surrounding material and modifies the local magnetic fields \citep{ferland09,ferriere11}. Since the magnetic field is generally coupled to the gas, its distribution is critically affected in this process. Furthermore, these dynamics can trigger new generations of stars by collect-and-collapse \citep{elmegreen77} or by radiation-driven implosion \citep{bertoldi89} processes. 

Magnetic fields are believed to play a crucial role in regulating star formation \citep{mestel56, mouschovias1999, mckee07, krumholz19}. However, it is unclear how the role of magnetic fields differs with conditions from quiescent molecular clouds to active environments such as H\,{\sc ii} regions. Stellar feedback of massive stars can compress molecular gas into dense fragments to form new massive stars \citep{deharveng05, tan14}, leading to questions regarding the significance of magnetic fields in triggered star forming conditions. Hence, a detailed study of magnetic field properties is necessary to better understand the dynamics of magnetic fields and how they relate to different star formation properties.

Several observations have revealed young star clusters associated with molecular clumps\footnote{In this paper, we use the term ``clump'' to describe a dense region within a cloud with a size of few parsecs and a mass of $\sim10^2 - 10^4 M_\odot$} around H\,{\sc ii} regions \citep{camargo11, thompson12, dewangan12}. The clumps are precursors to clusters \citep{battersby10}, and thus obtaining magnetic field properties in clumps is critical to assessing their stabilities. The efficiency of formation of star clusters and new stars can be constrained by considering two scenarios. Either the formation is regulated by strong magnetic fields, through the process of ambipolar diffusion \citep{mouschovias1991,mouschovias1999} or the turbulent gas motions, outflows, and stellar winds provide non-magnetic support against collapse \citep{mckee03,padoan2004, mackey11}. An observational test is to compare the relative strengths of the gravitational potential with the magnetic field using the mass-to-flux ratio \citep{crutcher04}. Additionally, if changes in magnetic strength have power-law dependency on column densities, it is possible to ascertain the influence of the magnetic field in channeling gas into dense regions.

Theoretical studies indicate the expansion of an ionized region can lead to more order in the magnetic field around the shell, while the molecular gas is driven to equipartition between thermal, magnetic, and turbulent energies \citep[][]{krumholz07,arthur11}. If the ionizing front is perpendicular to the ambient magnetic field, the expansion may be slowed in that direction. A few observational results \citep[e.g.,][]{tang09,santos14,chen17} have shown magnetic field orientations are aligned tangentially to ring-like structures and the field strength increases in the compressed neutral gas. 

While these studies give a hint of how magnetic fields may be affected by feedback, a systematic study was needed that ranges from parsec to sub-parsec scales and that connects turbulence, magnetic fields, and gravity. In this context, we explored the S235 complex using new polarization observations along with the distributions of dust and molecular gas. The paper is organized as follows. Section~\ref{sec2} gives an introduction to the S235 complex. Section~\ref{sec3} provides new near-infrared observational details and assembles relevant archival data products. Section~\ref{sec4} gives descriptions of the analyses and results. This includes selection of background stars, identification of dust clumps and estimation of the magnetic field properties. Section~\ref{sec5} discusses the implications of results and role of magnetic fields. Section~\ref{sec6} contains conclusions and a summary of the study. 

\section{The S235 complex} \label{sec2}

\begin{figure}[t!]
\epsscale{1.2}
\plotone{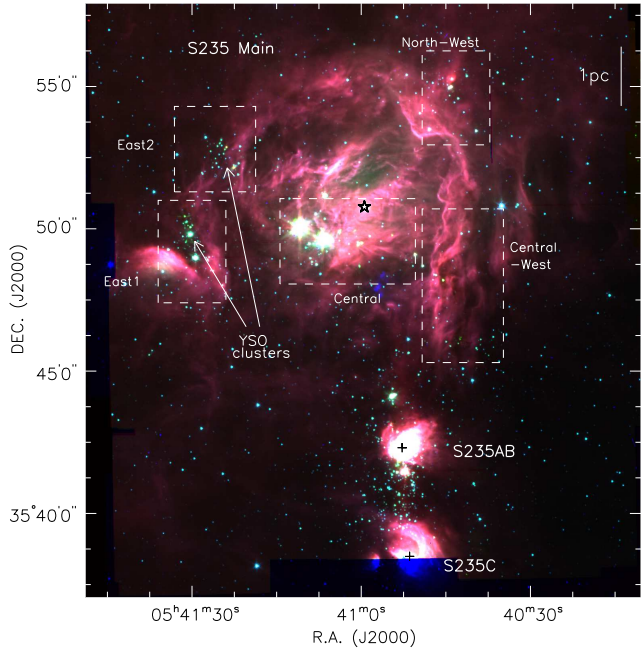}
\caption{Mid-infrared three-color image of the S235 complex, constructed using \textit{Spitzer} IRAC bands 8$\mu$m (red), 4.5$\mu$m (green) and 3.6$\mu$m (blue) \citep{fazio04}. The region spans $20\arcmin \times 21\arcmin$, corresponding to a projected physical scale of $9.56\,\mathrm{pc}\times 10.08\,\mathrm{pc}$ at a distance of $1.65\,\mathrm{kpc}$. Three separate H\,{\sc ii} regions: S235~Main, 235AB, and S235C are identified. The central ionizing O9.5V star for S235~Main is marked by the star symbol, while the ionizing B0.5V stars for S235AB and S235C are shown by cross symbol. Sub-regions with YSO clusters identified by \citet{kirsanova08} are shown by dashed boxes.}
\label{fig1}
\end{figure}

The \object{Sh2-235} (S235) star-forming complex is part of the giant molecular cloud G174+2.5 situated in the Milky Way Perseus spiral arm \citep{heyer96} and contains three known H\,{\sc ii} regions: `S235~Main' \citep{sharp59}, `S235AB', and `S235C'. The kinematic distance to S235 has been considered to be between $1.5\,\rm kpc$ and $1.8\,\rm kpc$ \citep{evans81,burns15}. We use a distance of $1.65_{-0.10}^{+0.12}\,\rm kpc$ based on the \textit{Gaia} DR2 \citep{brown18} parallax for the S235~Main ionizing star: \object{BD +35$^{\circ}$1201} of O9.5V type \citep{georgelin73}. Radial velocity measurements of photospheric lines in BD +35$^{\circ}$1201 give an LSR value of $-18\,\rm km\,s^{-1}$ \citep{kirsanova08}. Various studies of molecular emission from $^{12}$CO,$^{13}$CO, CS, NH$_{3}$ lines trace structures in the S235 complex over the velocity range $-25\,\rm km\,s^{-1}$ to $-15\,\rm km\,s^{-1}$  \citep{kirsanova08, kirsanova14,chavarria14}, indicating the star and molecular material are associated. \citet{dewangan16} used radio continuum emission at $1.4\,\rm GHz$ to reveal a spherical distribution of the ionizing radiation. {\it Spitzer} mid-infrared (MIR) images \citep{werner04} reveal a spherical shell-like morphology tracing the photo-dissociation regions (PDRs) (see Figure~\ref{fig1}). The ionizing star BD +35$^{\circ}$1201 appears in projection to be located approximately at the center of the shell. 

In addition, the S235AB and S235C regions (smaller, southern part of the complex, see Figure~\ref{fig1}) are powered by B0.5V stars \citep{felli97,bieging16}. Distance to S235AB has been obtained to be $1.56_{-0.08}^{+0.09}\,\rm kpc$ from water maser parallax measurements by \citet{burns15}. \citet{felli04} studied these regions with mm spectral and continuum observations along with far-infrared data, identifying the presence of a molecular core hosting a young stellar object (YSO) with a bipolar outflow. Using VLA radio cm observations and \textit{Spitzer} data, \citet{felli06} suggested that the YSO is the result of star formation triggered by the ionizing wind of the massive star in S235A. \citet{dewangan11} identified YSOs in S235~Main and S235AB using \textit{Spitzer} and near-infrared photometry. Subsequent studies \citep[see][]{kirsanova14,dewangan16} located these YSOs in association with the molecular gas along the shell and edges of the H\,{\sc ii} regions (see Figure~\ref{fig1}). All of these studies reveal that the S235 complex is dynamic and that triggered star formation is evident around the S235 Main region. However, there has been no study of the magnetic field properties in this complex. Hence, the S235 complex is an ideal laboratory to study the interaction of H\,{\sc ii} regions, molecular gas, and magnetic fields. 

The magnetic field orientations in the S235 complex can be probed using near-infrared (NIR) background starlight polarimetry. The dust grains in the molecular regions align their long axes perpendicular to the local magnetic field due to radiative torques \citep{laz07, and15}. Hence, the light from background stars passing through the dust becomes weakly polarized due to dichroic extinction \citep{hall49, hiltner49}. The resulting polarized starlight position angles are parallel to the plane-of-sky (POS) magnetic field orientations of the region.

\section{Observations and Data Sets} \label{sec3}

We obtained new polarimetric observations and combined these with archival multi-wavelength imaging and spectral data to study the magnetic field properties of the S235 complex. The angular size of the selected region is $20\arcmin \times 21\arcmin$, centered at $\alpha_{2000}$ = 05$^{h}$40$^{m}$58.9$^{s}$, $\delta_{2000}$ = +35$\degr$47$\arcmin$25.7$\arcsec$. This size covered all the three H\,{\sc ii} regions and corresponds to a physical scale of about $9.59\,\mathrm{pc}\times 10.08\,\mathrm{pc}$ at a distance of $1.65\,\mathrm{kpc}$.

\subsection{Near-infrared polarization observations} \label{NIRobs}

NIR linear polarimetric observations were obtained in $H$-band (1.6 $\mu$m) using two instruments: Mimir \citep{clemens07} and POLICAN \citep{car17, dev18b}. The Mimir instrument was located on the $1.8\,\mathrm{m}$ Perkins telescope outside of Flagstaff, AZ. Mimir used a 1024$\times$1024 pixel Aladdin III InSb detector array, operated at $33.5\,\mathrm{K}$. The instrument had a field-of-view (FOV) of 10$\times$10 arcmin$^2$ and pixel scale of $0.58\,\mathrm{arcsec}$. The POLICAN instrument operated on the $2.1\,\mathrm{m}$ OAGH telescope in Cananea, Sonora, Mexico. It had a 1024$\times$1024 pixel HgCdTe detector with a pixel scale of $0.32\,\mathrm{arcsec}$ that provided a FOV of 4$\times$4 arcmin$^2$. Both instruments measured linear polarization by the combination of a half-wave plate (HWP) as the modulator and a fixed wire-grid polarizer as the analyzer. Images obtained through different HWP orientation angles were used to determine the Stokes parameters $Q$ and $U$ for each measured source. 

\begin{figure}[!ht]
\epsscale{1.2}
\plotone{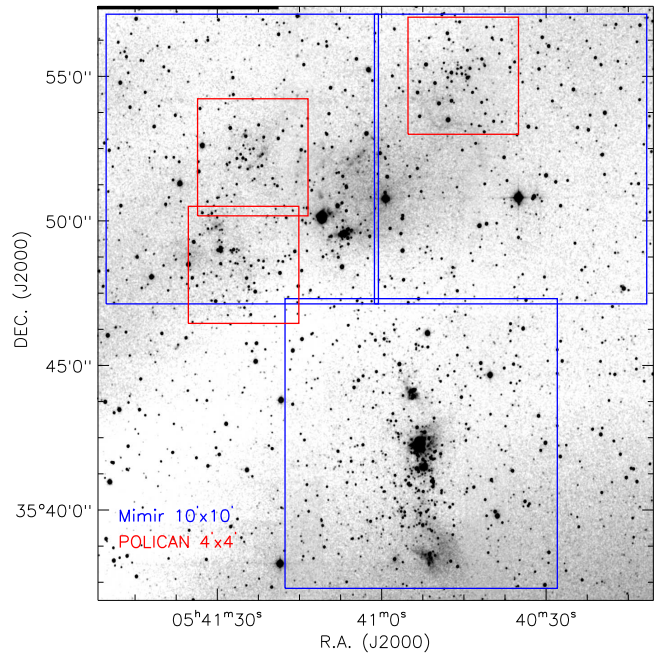}
 \caption{Near-infrared $H$-band image of the S235 complex obtained from 2MASS data. Colored boxes corresponding to the observed fields are shown. The blue boxes are $10\,\mathrm{arcmin}$ fields observed using Mimir. The red boxes are $4\,\mathrm{arcmin}$ fields observed using POLICAN.}
\label{fig2}
\end{figure}

Observations with Mimir were carried out in $H$-band during 2015 January for seven nights. Three target fields of $10\,\mathrm{arcmin}$ were chosen that covered the S235 region, with overlaps of $0.1\,\mathrm{arcmin}$ between adjacent fields. The field coverage is shown in Figure~\ref{fig2}. Each field was observed in six sky dither positions with images taken at 16 HWP angles. A total of 96 (6$\times$16) images were taken for each field, each with an exposure time of $15\,\mathrm{sec}$, for a total integration time of 24 minutes for each field. Each field was observed 14 times totaling 5.6 hours per field. The data reduction and calibration were carried out using the Mimir Software Package: Basic Data Processing (MSP-BDP), and the Photo POLarimetry tool (MSP-PPOL) yielding Stokes parameters $Q$, $U$, and polarization values for all the point sources in the fields. Details of the processes used in the Mimir software package are described in Galactic Plane Infrared Polarization Survey (GPIPS) articles \citep{clemens12a,clemens12b}.

Observations with the POLICAN instrument focused on reaching finer angular resolution and signal-to-noise ratios for regions with high stellar density. Three sub-fields of 4$\times$4 arcmin$^2$ were observed in $H$-band during 2017 September. The field coverage is shown in Figure~\ref{fig2}. Each set of observations contained 18 sky dither positions at each of 4 HWP angles $(0^{\circ}, 22$.$5^{\circ}, 45^{\circ}, 67$.$5^{\circ})$. Exposure times were $100\,\mathrm{sec}$, for total integration time of $2\,\mathrm{hours}$ for the 72 images (18$\times$4) per field. Basic reduction consisted of linearity correction, dark subtraction, flat-fielding and sky removal. Images were aligned and combined for each HWP angle \citep[see][]{dev18a}. They were then astrometry-corrected using data from 2MASS \citep{skrutskie06}. Aperture photometry was performed on each source in each of the HWP images. The resulting source intensities were used to compute the Stokes parameters. Descriptions of calibration procedures and polarimetric analyses are given in \citet{dev18b}. 

The polarization percentage ($P$) and equatorial position angle ($PA$) were computed from Stokes parameters $Q$, and $U$, for both instruments, as follows:
\begin{equation}
P = \sqrt{Q^{2}+U^{2}}\,, \label{eqn1}
\end{equation}
\begin{equation}
PA = \frac{1}{2}\tan^{-1}\left(\frac{U}{Q}\right). \label{eqn2}
\end{equation}

The polarization angles are measured from North-up toward East-left direction. The de-biased degree of polarization ($P^{'}$) was estimated using Ricean correction as prescribed in \citet{wardle74}:
\begin{equation}
P^{'} = \sqrt{P^{2} - \sigma_{P}^{2}}\,, \label{eqn3}
\end{equation}
where $\sigma_{P}$ is the polarization uncertainty computed by standard error propagation of the corresponding Stokes values and uncertainties \citep[see Appendix in][]{dev18b}.

The final results from the observations were data files containing coordinates, brightness magnitudes, Stokes parameters, polarization values, and corresponding uncertainties. Results from both instruments were combined to produce a single set of stellar polarization catalog. Sources observed multiple times from overlapped regions, and on both instruments, were checked for repeatability.

\begin{figure}[!ht]
\epsscale{1.2}
\plotone{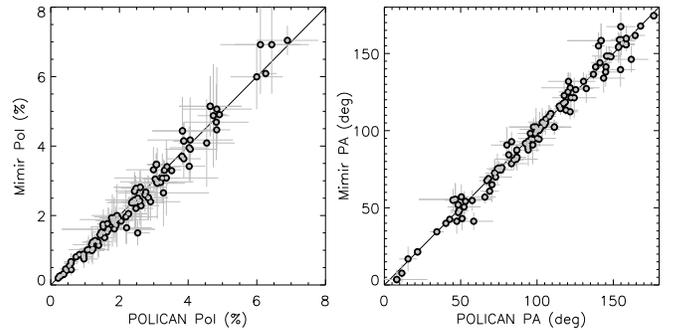}
 \caption{Comparison of Mimir and POLICAN polarization for stars repeated in both fields, having $\sigma_{P}<3\%$. The \textit{left panel} compares degree of polarization and the \textit{right panel} compares the position angles. In each panel, the solid black line represents equality with unit slope and zero offset.}
\label{fig3}
\end{figure}

We found that 215 stars were in common between the Mimir and POLICAN fields. Figure~\ref{fig3} compares the polarization values for repeated stars having $\sigma_{P}<3\%$. The plot shows excellent correlation with values distributed around the equality line. The mean polarization percentage difference of the two sets was $0.20\pm0.06\%$ with standard deviation of $0.4\pm0.1\%$. The mean position angle difference was $0.6\pm0.4^{\circ}$ with dispersion of $3.8\pm0.7^{\circ}$. In order to retain only one value of each repeated star for the final data set, we chose the value having the highest polarization signal-to-noise ($P_{SNR}=P^{'}/\sigma_{P}$) between both measurements. The selected stars were added to the combined set containing unique Mimir and unique POLICAN stars. To favor the most reliable polarimetric data, we retained only those sources meeting the criteria $P_{SNR}>1$ and $\sigma_{P}<3\%$. This resulted in 623 sources with reliable stellar polarization values. 

Table~\ref{tab1} lists the polarimetric properties of the 623 stars in the S235 field. The table columns are ordered as follows: star ID, equatorial coordinates RA and DEC, Mimir or POLICAN $H$-band magnitudes, de-biased polarization percentage, equatorial position angle, and the equatorial Stokes $Q$ and $U$ parameters. The uncertainties are given in parentheses.

\begin{table*}
\movetabledown=13mm
\footnotesize
\centering
\renewcommand{\arraystretch}{1.2}
\setlength{\tabcolsep}{0.15in}
\caption{Polarimetric properties of the 623 stars in the S235 complex.}
\label{tab1}
\hspace*{-2cm}\begin{tabular}{cccccccc}
\hline 
\hline
 ID & R.A. & DEC. & $H$-band & $P^{'}$ & $PA$ & $Q$ & $U$ \\
  & (J2000) &  (J2000) & (mag) & (\%)  & (deg) & (\%) & (\%) \\
\hline

1 & 85.10488 & 35.90900 &  11.83 (0.01) & 0.4 (0.2) & 142 (12) & 0.1 (0.1)   & $-$0.4 (0.2) \\
2 & 85.10931 & 35.89397 &  15.12 (0.03) & 6.2 (2.2) & 159 (10) & 5.0 (2.3)   & $-$4.2 (2.2) \\
3 & 85.11411 & 35.91162 &  13.58 (0.03) & 1.0 (0.6) & 10 (18)  & 1.2 (0.6)   & 0.4 (0.6) \\
4 & 85.11484 & 35.84210 &  12.25 (0.01) & 1.1 (0.2) & 130 (5)  & $-$0.2 (0.1)& $-$1.0 (0.2) \\
5 & 85.11559 & 35.87341 &  15.09 (0.03) & 3.2 (2.4) & 133 (21) & $-$0.3 (2.3)& $-$4.0 (2.4) \\
6 & 85.12303 & 35.73406 &  14.93 (0.02) & 4.1 (2.0) & 132 (14) & $-$0.4 (2.0)& $-$4.5 (2.0) \\
7 & 85.12359 & 35.93396 &  11.94 (0.01) & 0.4 (0.2) & 9 (12)   & 0.5 (0.2) & 0.1 (0.2) \\
8 & 85.12410 & 35.64257 &  12.28 (0.01) & 0.7 (0.6) & 131 (26) & $-$0.1 (0.6)& $-$1.0 (0.6) \\
\hline          
\end{tabular}
\begin{flushleft}
\tablecomments{Only few rows of the entire table are shown here. The complete list is available online in machine-readable form. Uncertainties are given in parentheses.}
\end{flushleft}
\end{table*}

\subsection{\textit{Planck} polarized dust emission}

The {\it Planck} all sky survey data \citep{planck16a} were used to obtain a map of polarized dust emission towards the S235 complex. We retrieved the Intensity $I$, Stokes $Q$, and $U$ sky maps observed with the High Frequency Instrument (HFI) at $353\,\rm GHz$. The procedures of map-making, calibration, and corrections are described in \citet{planck16b}. We smoothed the maps from an initial angular resolution of 5$\arcmin$ to 6$\arcmin$ to increase the signal-to-noise ratio and to minimize beam depolarization effects. The intensity and polarization maps were converted from the $K_\mathrm{CMB}$ temperature scale to MJy sr$^{-1}$ using the unit conversion and color correction factor of 246.54. The intensity map was also subtracted for the Galactic zero level offset of 0.0885 MJy sr$^{-1}$ \citep{planck16b}. The corrected maps of Stokes parameters were analyzed using the surface brightness of each pixel to calculate the polarization values, as given in Equation~\ref{eqn1}~and~\ref{eqn2}. This resulted in linear polarization map of dust emission integrated along the line-of-sight (LOS). More details on polarization analysis is presented in \citet{planck16c}. The electric field polarization position angles were rotated by 90$^{\circ}$ to yield POS magnetic field orientations. The rotated polarization angles are represented in Equatorial coordinates and measured from North-up toward East-left direction.
 
\begin{table*}[!ht]
%\scriptsize
\setlength{\tabcolsep}{0.01in}
\centering
\caption{Archival data sets adopted in this work.}
\label{tab2}
\vspace{-0.4cm}
\begin{tabular}{lcccccr}
\hline 
\hline
Survey 														 	& Wavelength(s)       						& Resolution     							& Reference\\
\hline
UKIRT Infrared Deep Sky Survey (UKIDSS)                & $0.88-2.2 \mu$m  						& $\sim0.8\arcsec$          									&\citet{lawrence2007}\\
Two Micron All Sky Survey (2MASS)                            	& $1.25-2.2 \mu$m  						& $\sim$2.5$\arcsec$          									&\citet{skrutskie06}\\
{\it Spitzer} IRAC GTO program ID:201									& 3.6, 4.5, 5.8, 8.0  $\mu$m             	& $\sim2\arcsec$						          			&\citet{fazio04}\\  

{\it Planck} dust polarization data                  			& 850 $\mu$m 									& $\sim5\arcmin$        										&\citet{planck16a}\\

Bolocam Survey extended to outer Galaxy        					&$1.1\,\mathrm{mm}$                     & ${\sim}$33$\arcsec$						        	&\citet{ginsburg13}\\

Extended Outer Galaxy Survey $^{13}$CO (J = 1--0)       	& 2.7 mm	 									& $\sim45\arcsec$; velocity 0.21~km\,s$^{-1}$       			&\citet{brunt04}\\
\hline          
\end{tabular}
\end{table*}

\subsection{Archival data sets}  
  
Additional archival data sets, ranging from NIR to radio wavelengths, were used in our study. NIR photometry was obtained from 2MASS and UKIDSS data to estimate visual extinction for polarization stars. MIR images, which reveal PDRs were obtained from \textit{Spitzer} IRAC guaranteed time observations (GTO) \citep{fazio04}. Dust continuum emission at $1.1\,\mathrm{mm}$ was obtained from the Bolocam Galactic Plane Survey, version 2.1 \citep{aguirre11, ginsburg13} with extension to the outer Galaxy. These data were used to identify molecular clumps and estimate H$_{2}$ column densities. $^{13}$CO molecular line data ($\rm J=1-0$, $110.2\,\rm GHz$) with rms sensitivity of $\sim0.13\,\rm K$ \citep{jackson06}, were obtained from the Extended Outer Galaxy Survey \citep{brunt04} taken using the FCRAO telescope. The molecular data were used to find gas velocity properties. Table~\ref{tab2} summarizes the survey wavelengths and resolutions of the data sets.

\section{Analyses} \label{sec4}

\subsection{Selection of Background stars}

Since the magnetic field inside the S235 complex was to be probed using the polarimetry of background stars, it is important to identify foreground and background stars among all the 623 sources observed. 
The selection of background stars involved three steps. Step 1: Identification and exclusion of young stellar objects using the NIR color-color diagram. Step 2: Use of \textit{Gaia} distances to classify foreground and background stars for the stars with \textit{Gaia} matches. Step 3: Stars without \textit{Gaia} distances had their extinctions analyzed to discern their foreground or background natures.

 \begin{figure*}[!ht]
\epsscale{0.85}
 \plotone{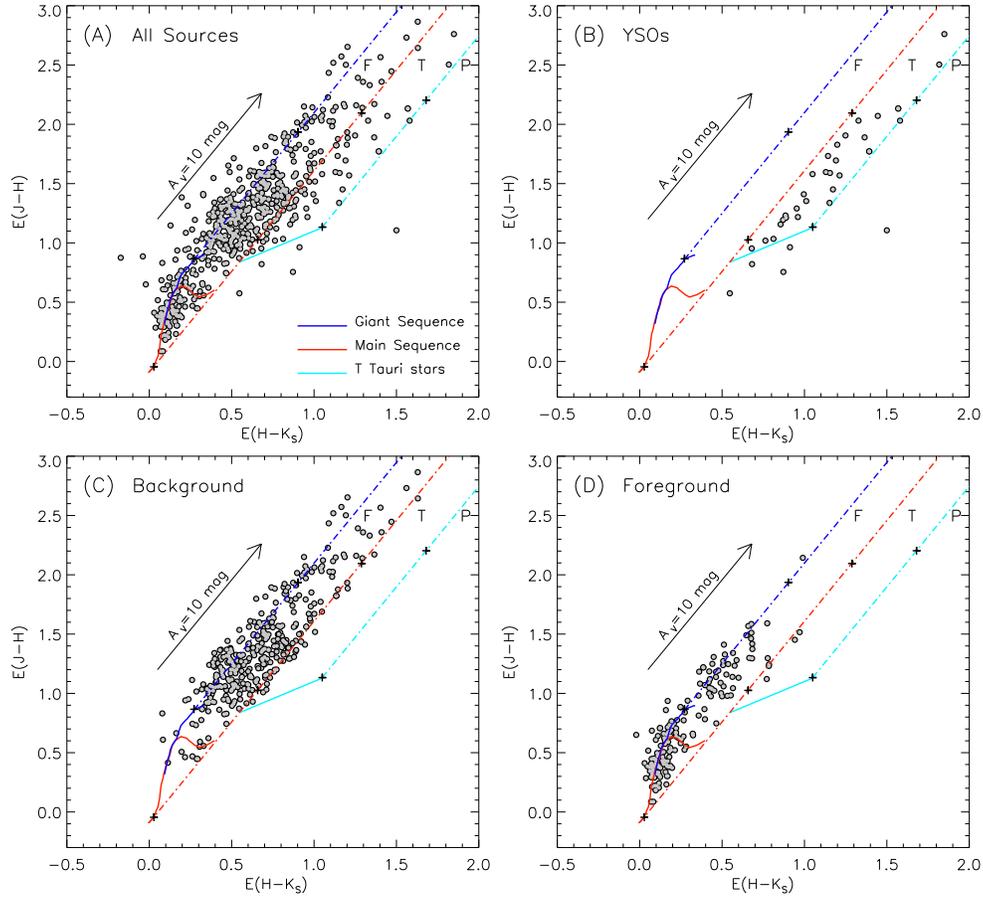}
 \caption{NIR color-color diagram $(J-H)$ vs $(H-K)$ for the 612 polarimetric sources in the S235 complex. The solid curves represent the unreddened locus of main sequence (blue), giants (red), and T Tauri stars (cyan). The loci are extended as three parallel dashed lines to represent reddening. An extinction arrow of $10\,\rm{mag}$ is shown for reference. The diagrams are classified into three regions: `F,' `T,' and `P' (see text). \textit{Panel}(A) shows distribution of all the 612 polarimetric sources that matched to the 2MASS and UKIDSS catalog. \textit{Panel}(B) shows the colors of the 38 YSOs falling within the `T' and `P' regions. \textit{Panel}(C) shows the colors of the 375 background stars identified using \textit{Gaia} distances and extinction. \textit{Panel}(D) shows the colors of the 185 foreground stars from the remaining sample.}
\label{fig4}
\end{figure*}

\subsubsection{Identification of YSOs}

The reddening and polarization of stars is related to the extinction by dust in clouds and in the diffuse interstellar medium \citep{martin92,whittet92}. However, young stellar objects and NIR-excess sources exhibit polarization due to reflected/scattered light from their circumstellar disks/envelopes and surrounding nebulosity \citep{bastien96,tamura06}. Hence, their polarizations do not trace magnetic field directions and such sources need to be excluded. 

\citet{dewangan11} and \citet{chavarria14} presented catalog of YSOs in the S235 complex identified using NIR and Spitzer data. Their study lists different classes of YSOs, but they do not properly account for YSO contaminations. Examination of their catalog shows many of the YSOs reveal NIR colors resembling giant and main sequence stars. Hence, we did not use these catalogs for YSO identification and performed separate analysis using NIR data.

The NIR color-color diagram provides a useful technique for identifying populations of YSOs \citep{lada92}. Using the unreddened loci for main sequence, giants, and young stars we can classify sources based on their color excess. The loci can be extended to account for reddening using the NIR extinction law \citep{mathis90}. Within the $(J-H)$ vs $(H-K)$ color-color diagrams, the distributions of sources typically fall into three regions, designated here as `F,' `T,' and `P' \citep[see][]{sugitani02,mallick12}. The `F' region contains field stars that include main-sequence, giants, and Class III (diskless) YSOs. The `T' region contains classical T Tauri stars (Class II protostars). The `P' region contains Class I protostars and Herbig Ae/Be stars with large NIR excess.

The 623 polarimetric stars were matched to the 2MASS point source catalog \citep{skrutskie06} to yield 559 stars with $J$, $H$, and $K$-bands photometry. The stars with no 2MASS magnitudes were matched with UKIDSS catalog \citep{lawrence2007} to obtain an additional 53 stars with NIR photometry. The UKIDSS magnitudes were calibrated to 2MASS system as described in \citet{hodgkin09}. The combination of 2MASS and UKIDSS matched sources resulted in total of 612 polarimetric stars having magnitudes in all the three $J$, $H$, and $K$-bands. Figure~\ref{fig4}(A) shows the $E(J-H)$ vs $E(H-K)$ color-color diagram for this sample of 612 stars. The loci of curves for main sequence (MS) stars and giants were taken from \citet{bessell88} and converted to the 2MASS system using equations from \citet{carpenter01}. The classical T Tauri (CTTS) loci was taken from \citet{meyer97}. These curves were then extended using reddening vectors drawn from the tip (spectral type M4) of the giant branch (blue line), from the base (spectral type A0) of the MS branch (red line), and from the tip of the intrinsic CTTS line (cyan line) using extinction ratios $A_{J}/A_{V}=0.265,A_{H}/A_{V}=0.155,$ and $A_{K}/A_{V}=0.09$, obtained from \citet{cohen81}.

In Figure~\ref{fig4}(A), 38 sources were found to fall within the `T' and `P' regions. These sources were classified as YSOs and are shown separately in Figure~\ref{fig4}(B). There were also 14 sources having color excesses not falling into any classification. These sources, along with the YSOs, were excluded from the later analyses.

The remaining sample of 560 stars fell in the `F' region, indicating that they are most likely normal main sequence and giant stars. These stars should be free from intrinsic polarization and their observed polarizations are due to dichroic extinction. The 560 sources were retained for the study and include foreground stars and background stars to the S235 complex.

\begin{figure}[!ht]
\epsscale{1.2}
 \plotone{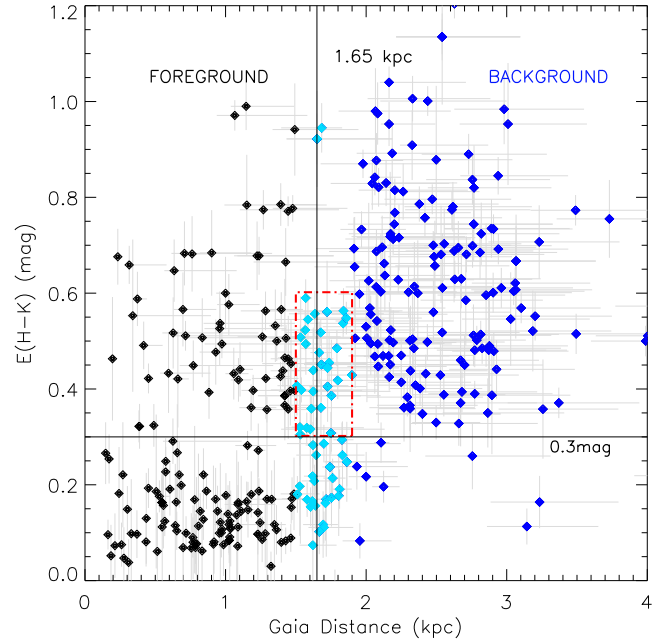}
 \caption{Plot of color excess $E(H-K)$ against distances for the 382 \textit{Gaia} matched stars. The solid black vertical line represents the $1.65\,\rm kpc$ distance to the S235 complex. The solid horizontal line shows a color excess of $0.3\,\rm mag$. The foreground stars are shown as black symbols. The background stars are shown as dark blue symbols. The stars associated with the S235 complex, ranging from $1.5-1.9\,\rm kpc$ distances, are shown as cyan symbols. A red dashed box covering $0.3<E(H-K)<0.6$ is shown to represent the region containing embedded stars in the S235 complex.}
\label{fig5}
\end{figure}

\subsubsection{\textit{Gaia} distances}

The \textit{Gaia} collaboration DR2 provides parallax values for sources as faint as $20\,\rm mag$ in $G$-band. \citet{bailer18} provided a catalog of distance estimate for the sources in \textit{Gaia} DR2. Their inferred distances are the modes of the posterior probability density functions from the measured parallax values, obtained using a varying distance prior. From the \citet{bailer18} catalog, we retrieved distances and astrometric information for the sources within the S235 field. We cross-matched the 560 field stars to \textit{Gaia} data, using a match radius of $2\,\mathrm{arcsec}$. In total, 382 stars from the sample were found to have \textit{Gaia} matches. 

Figure~\ref{fig5} plots color excess against distance for the 382 \textit{Gaia} matched stars. The solid black vertical line represents the $1.65\,\rm kpc$ distance to the S235 complex. The stars show a positive correlation of color excess with distance. The distribution has a step-wise trend with the step centered near the distance of $1.65\,\rm kpc$. A horizontal solid black line of $E(H-K)=0.3$ is shown to represent the average of the step, distinguishing regions of low and high color excess. The low color excess stars mostly have distances less than $1.65\,\rm kpc$. Stars with $E(H-K)>0.3$ mainly show distances greater than $1.65\,\rm kpc$.

The stellar distribution in Figure~\ref{fig5} was interpreted to contain three distinct populations of stars, namely: foreground, background, and stars associated with, or embedded in the S235 complex. The foreground stars were considered those with distances less than $1.5\,\rm kpc$. Of the 382 stars, 182 stars fell in this category and are shown as black symbols in Figure~\ref{fig5}. The background stars were those having distances greater than $1.9\,\rm kpc$. A total of 160 stars met this criterion and are shown as dark blue symbols in Figure~\ref{fig5}. The background stars predominantly have color excess of $E(H-K)>0.3$, indicating they are highly extincted.

The stars associated with the S235 complex are the sample located near the distance of $1.65\,\rm kpc$, with a range of $1.5-1.9\,\rm kpc$. Although this distance range is much larger than typical cloud sizes, we consider the stars in it to be associated with the S235 complex given the distance uncertainties of the \textit{Gaia} DR2 data. The stars that fell in the associated region are shown as cyan symbols in Figure~\ref{fig5}. This sample includes stars that lie outside the S235 complex and stars that are embedded in it. The embedded stars are likely the subset that are moderately extincted, with color excesses of $0.3<E(H-K)<0.6$. There were 40 stars satisfying this condition and were deemed embedded. A red dashed box in Figure~\ref{fig5} is shown to represents the region containing the embedded stars. The 160 background stars and the 40 embedded stars formed a total of 200 \textit{Gaia} matched stars for use in the magnetic field analyses.

\begin{figure*}[!ht]
\epsscale{1.0}
\plotone{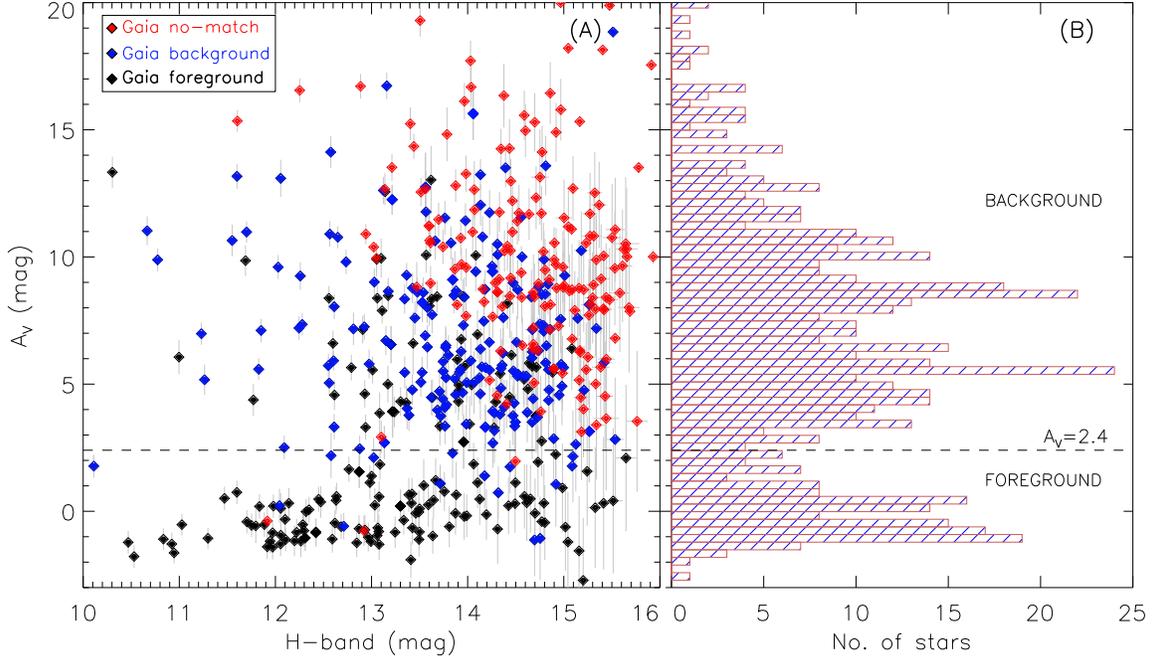} 
 \caption{\textit{Panel}(A) compares extinction values for the 560 field stars against their $H$-band magnitudes. The \textit{Gaia} matched embedded and background stars are shown in blue symbols, whereas the \textit{Gaia} matched foreground stars are shown in black symbols. The \textit{Gaia} un-matched stars are shown in red symbols. \textit{Panel}(B) shows histogram distribution of the same extinction values. In both the panels the dashed horizontal line represents $A_{V}=2.4\,\rm mag$, which corresponds to $E(H-K)=0.3\,\rm{mag}$ as shown in Figure~\ref{fig5}.}
   \label{fig6}
\end{figure*}

\subsubsection{Stellar extinction}

The 178 field stars that did not have \textit{Gaia} matches were examined further to classify their locations based on their extinction values. Since extinction is attributed to the amount of intervening dust in the LOS of a star, it can be used to classify stars as being either foreground or background to a cloud \citep{santos17, hoq17}.

The stellar extinction values can be calculated using the color excess obtained from 2MASS and UKIDSS magnitudes. The ratio of total to selective extinction $R$, is given as $R_{XY}=A_{V}/E(X-Y)$. The value of ${R}$ in the NIR are $R_{HK}=15.87, R_{JH}=9.35, R_{JK}=5.89$ \citep{rieke85}. Taking into account the intrinsic colors (mean intrinsic color was determined to be $\sim$0.15 from a nearby control field), the relation for visual extinction to NIR color excess is written as:
\begin{equation} \label{Av}
A_{V} = 15.87 \times [(H-K_{s}) - 0.15]. 
\end{equation} 

The estimated extinction values for all the 560 field stars are plotted against their $H$-band magnitudes in Figure~\ref{fig6}(A). The \textit{Gaia} matched embedded and background stars are shown in blue symbols, whereas the \textit{Gaia} matched foreground stars are shown in black symbols. The 178 \textit{Gaia} un-matched stars are shown in red symbols. Most of the \textit{Gaia} matched stars are distributed across both low and high extinction zones but favoring brighter $H$-band magnitudes. The \textit{Gaia} un-matched stars are found predominantly at high extinction zones and at fainter $H$-band magnitudes.

\citet{clemens20} compared extinction values for GPIPS stars with and without \textit{Gaia} distances. They showed that the GPIPS stars not matching to the \textit{Gaia} sample were the most extincted of all, reaching $A_{V}$ values up to $30\,\rm mag$. To determine which of the \textit{Gaia} un-matched stars are background to the S235 complex, we compared their extinctions with the properties of \textit{Gaia} matched background stars.

In Figure~\ref{fig5}, the \textit{Gaia} background stars predominantly show a color excess of $E(H-K)>0.3$, or $A_{V}>2.4\,\rm mag$, using Equation~\ref{Av}. A dashed horizontal line is shown in Figure~\ref{fig6} to represent $A_{V}=2.4\,\rm mag$. More than $97\%$ of \textit{Gaia} un-matched stars have $A_{V}>2.4\,\rm mag$, indicating they are mostly likely embedded or background stars. 

A histogram of $A_{V}$ values is shown in Figure~\ref{fig6}(B) and reveals two distinct populations of stars. There is an initial increase in extinction seen as a small bump up to $A_{V}=2.4\,\rm mag$. This contribution is due to the dust in the interstellar medium being revealed by foreground stars. Beyond $A_{V}=2.4\,\rm mag$, there is rise in $A_{V}$ due to extinction by dust in the cloud. Based on these results and from comparison with Figure~\ref{fig5}, we infer that \textit{Gaia} un-matched stars having $A_{V}>2.4\,\rm mag$ are embedded or behind the cloud. This analysis resulted in 175 \textit{Gaia} un-matched stars being classified as embedded or background. These 175 stars were merged with the \textit{Gaia} matched embedded and background stars (200 stars) to obtain a total of 375 stars suitable for magnetic field analyses. The 375 stars are shown on NIR color-color diagram in Figure~\ref{fig4}(C).

\begin{figure*}[!ht]
\epsscale{0.85}
 \plotone{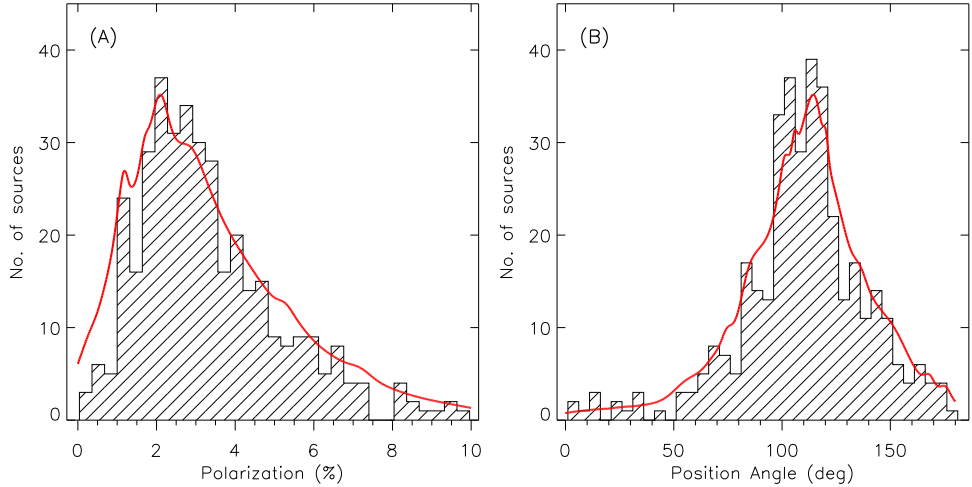}
 \caption{Histograms of polarization values for the 375 foreground-corrected embedded and background stars. \textit{Panel}(A) shows distribution of de-biased polarization percentage. The mean polarization value for the sample of stars was $2.5\pm0.2\%$ with a standard deviation of $1.4\pm0.1\%$. \textit{Panel}(B) shows the polarization position angles. The mean position angle was $114\pm4^{\circ}$ with a dispersion of $22\pm1^{\circ}$. The red solid line in each panel represents the Gaussian probability density distribution for the polarization values, considering the uncertainties, and placed into corresponding histogram bins.}
\label{fig7}
\end{figure*}

\subsubsection{Accounting for Foreground Polarization}

Since the dust foreground to the S235 complex contributes its polarization signal to that of every background star, those contributions should be removed from the background star polarizations prior to magnetic field analyses. We assumed a uniform layer of interstellar dust in front of the S235 complex. This assumption is reasonable as there is only one molecular velocity component along the LOS (the S235 component; see Section~\ref{COcloud}) and there is no other cloud either foreground or background to the S235 complex.

Based on the previous steps of embedded and background star selection, the remaining 185 field stars (182 \textit{Gaia} matched and 3 \textit{Gaia} un-matched) were classified as foreground, and are shown in the NIR color-color diagram in Figure~\ref{fig4}(D). Of these 185 stars, we selected only the sub-sample having a low color excess of $E(H-K)<0.3$, to remove any bias from high extinction foreground stars. We find the average polarization of this foreground sub-sample to be $\sim0.3\%$. Using the individual Stokes values of foreground stars, we estimated the weighted mean Stokes value of the foreground contribution. The values are $Q_{fg}=0.19\pm0.04\%$ and $U_{fg}=-0.11\pm0.03\%$. These were subtracted from the Stokes values of background stars to produce foreground-corrected Stokes values as $Q_{cor}=Q-Q_{fg}$ and $U_{cor}=U-U_{fg}$. The de-biased degree of polarization and position angles were then calculated using Equations \ref{eqn1} to \ref{eqn3}.

\begin{table*}
\movetabledown=13mm
\scriptsize
\centering
\renewcommand{\arraystretch}{1.2}
\setlength{\tabcolsep}{0.065in}
\caption{Polarimetric and photometric properties of the 375 foreground-corrected embedded and background stars in the S235 complex.}
\label{tab3}
\hspace*{-2cm}\begin{tabular}{cccccccccccc}
\hline 
\hline
 ID & R.A. & DEC. & $P^{'}$ & $PA$ & $Q$ & $U$ & $J$ \tablenotemark{\scriptsize{$\rm1$}} & $H$ & $K$ & $A_{V}$ & Distance \tablenotemark{\scriptsize{$\rm2$}} \\
  & (J2000) &  (J2000) & (\%)  & (deg) & (\%) & (\%) & (mag) & (mag)  & (mag) & (mag) & (kpc) \\
\hline

 1 & 85.10488 & 35.90900 & 0.3 (0.2) & 129 (12) & $-$0.1 (0.1) & $-$0.3 (0.2) & 13.22 (0.02) & 11.85 (0.02) & 11.25 (0.02) & 7.1 (0.5) & $1.9 \rvert_{1.3}^{3.2}$ \\
 2 & 85.10932 & 35.89397 & 5.9 (2.3) & 159 (10) &    4.8 (2.3) & $-$4.1 (2.2) & 16.38 (0.09) & 15.34 (0.08) & 14.74 (0.09) & 7.2 (1.9) & $2.1 \rvert_{0.9}^{4.2}$ \\ 
 3 & 85.11484 & 35.84210 & 1.0 (0.2) & 124 (5)  & $-$0.4 (0.2) & $-$0.9 (0.2) & 13.67 (0.02) & 12.25 (0.02) & 11.64 (0.02) & 7.2 (0.5) & $3.0 \rvert_{1.9}^{5.0}$ \\ 
 4 & 85.11559 & 35.87341 & 3.1 (2.4) & 131 (21) & $-$0.4 (2.3) & $-$3.9 (2.4) & 16.10 (0.11) & 14.90 (0.07) & 14.12 (0.09) & 9.9 (1.2) &  ... \\
 5 & 85.12303 & 35.73406 & 4.0 (2.0) & 130 (14) & $-$0.6 (2.0) & $-$4.4 (2.0) & 16.32 (0.19) & 15.05 (0.08) & 14.36 (0.08) & 8.6 (1.4) &  ... \\      
 6 & 85.12410 & 35.64257 & 0.6 (0.6) & 125 (26) & $-$0.3 (0.6) & $-$0.8 (0.6) & 13.89 (0.02) & 12.28 (0.02) & 11.66 (0.02) & 7.3 (0.4) & $2.0 \rvert_{1.2}^{4.0}$ \\
 7 & 85.12731 & 35.80114 & 1.5 (0.4) & 56 (9)   & $-$0.6 (0.4) &    1.4 (0.4) & 14.59 (0.03) & 13.03 (0.02) & 12.31 (0.02) & 9.0 (0.5) & $2.2 \rvert_{1.1}^{4.3}$ \\
 8 & 85.12906 & 35.84469 & 2.2 (1.5) & 93 (22)  & $-$2.6 (1.5) & $-$0.3 (0.5) & 15.55 (0.05) & 14.26 (0.04) & 13.86 (0.04) & 3.9 (0.9) & $1.6 \rvert_{0.7}^{3.7}$ \\

\hline          
\end{tabular}
\begin{flushleft}
\tablecomments{Only few rows of the entire table are shown here. The complete list is available online in machine-readable form. Missing data are shown here as ellipses but are represented by the number 0.0 in the online form. Uncertainties are given in parentheses.\\
$^1$ $J$, $H$, and $K$ band magnitudes from 2MASS or UKIDSS data.\\
$^2$ Distance values from \citet{bailer18} catalog with their low and high uncertainties.}
\end{flushleft}
\end{table*}

\begin{figure*}
\epsscale{0.82}
 \plotone{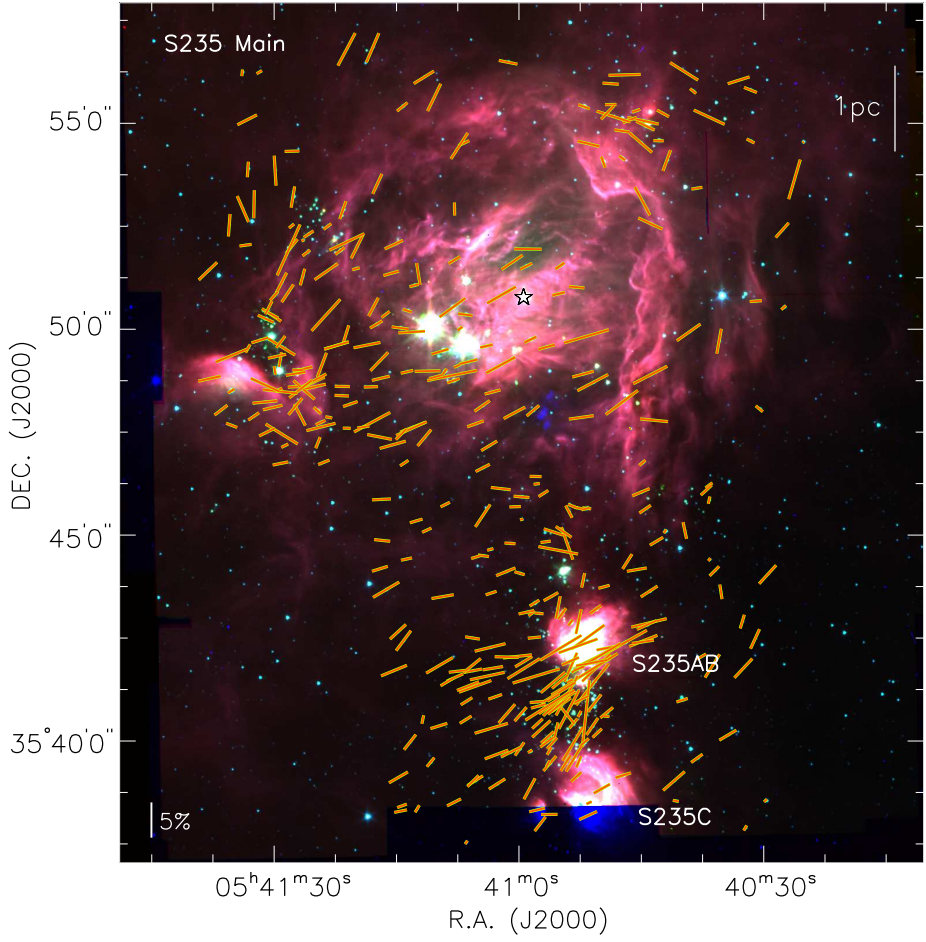}
 \caption{Magnetic field morphology in the S235 complex. The background image is the same \textit{Spitzer} IRAC three-color-composite map shown in Figure~\ref{fig1}. The NIR background starlight polarizations are indicated by red-yellow pseudo vectors. The lengths of the vectors indicate the polarization percentages and a reference vector length is shown at the bottom left. The position angle of these vectors trace the POS magnetic field orientations in the S235 complex.}
\label{fig8}
\end{figure*}

The distribution of 375 foreground-corrected embedded and background star polarizations are shown as histograms in Figure~\ref{fig7}. To represent the histograms accurately and to account for value uncertainties, each star value and its uncertainty were replaced by Gaussian probability densities, accumulated and placed into the corresponding histogram bins. The red solid lines in Figure~\ref{fig7} represent the net Gaussian probability density distribution for the polarization values. The degree of polarization for the sample ranges between 1 to 10$\%$ with a mean of $2.5\pm0.2\%$ and a standard deviation of $1.4\pm0.1\%$. The mean polarization value is similar to the typical values of NIR dust polarization in molecular clouds \citep{chapman11, clemens20} and larger than the average interstellar polarization of $\sim0.5\%$ in $H$-band \citep{martin92,whittet92}. About 25\% of the S235 stars exhibit polarization percentage greater than $4\%$ and so could be probing more extincted regions of the cloud. The polarization efficiency in the S235 complex is discussed in Section~\ref{PE}. The polarization angles have a mean value of $114^{\circ}$ and a dispersion of about $22^{\circ}$. These final polarization values represent the POS magnetic field orientations and are shown in Figure~\ref{fig8} as pseudo vectors on the \textit{Spitzer} mid-infrared image of the S235 complex.

Table~\ref{tab3} lists the polarimetric and photometric properties of the 375 foreground-corrected embedded and background stars. The table columns are ordered as follows: star ID, equatorial coordinates RA and DEC, de-biased polarization percentage, equatorial position angle, 2MASS or UKIDSS magnitudes in $J$, $H$, and $K$-bands, visual extinction, and \textit{Gaia} DR2 distances from \citet{bailer18} with their low and high values.

\subsection{Clump Identification and Clump properties}

%Previous studies from \citet{dewangan16} presented list of molecular clumps based on results from Bolocam catalog \citet{ginsburg13}. These clumps were extracted using Bolocat, that included faint nearby emission and filamentary structures. Our goal is to trace magnetic field information around massive clumps (size of few sub-parsecs). Hence it is important to identify dense regions with significant peaks and obtain clear clump boundaries without any overlaps. To improve on the existing clump boundaries, we re-ran the Bolocat algorithm with new detection parameters

\begin{figure}[!ht]
\epsscale{1.2}
 \plotone{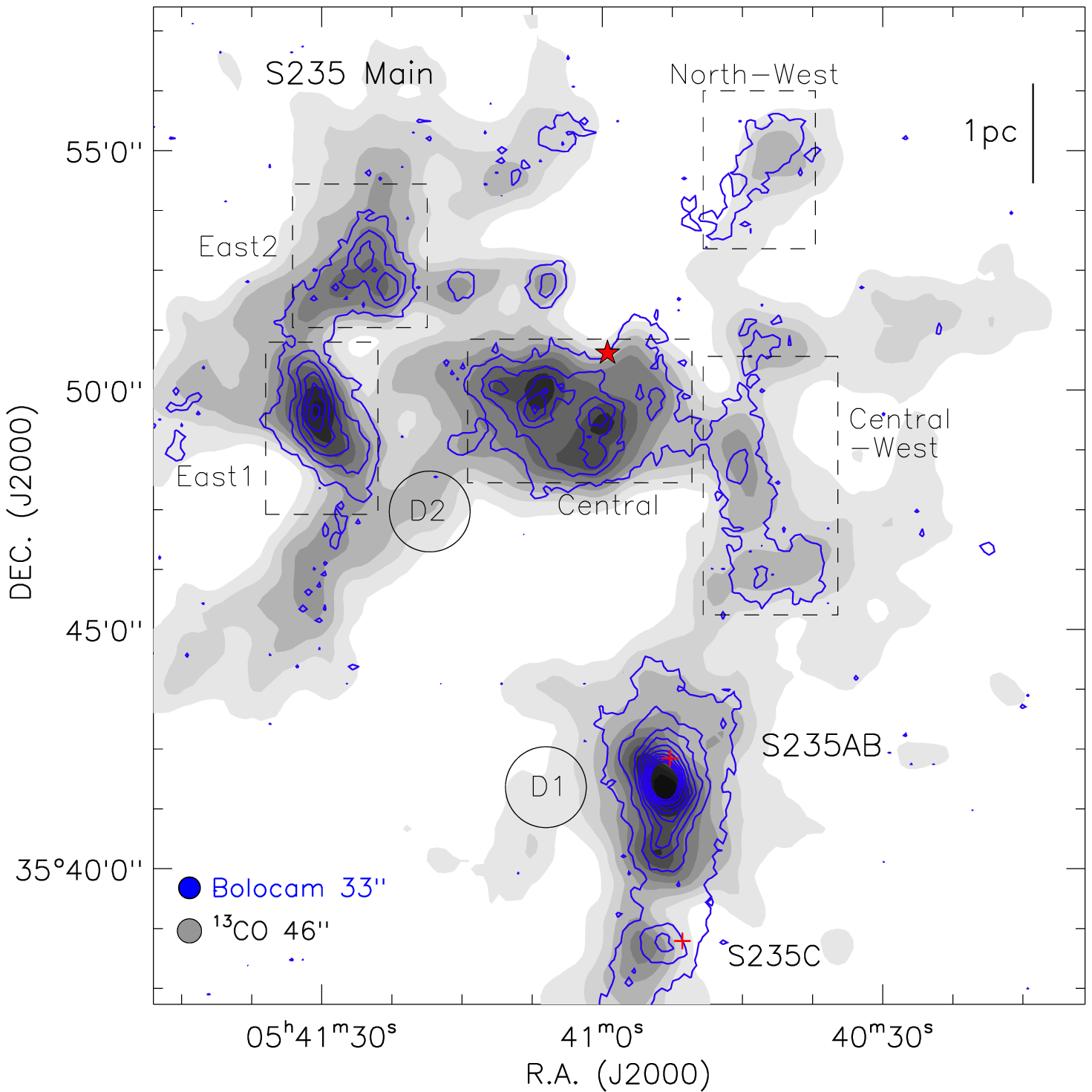}
 \caption{The distribution of molecular gas and dust emission in the S235 complex. The $^{13}$CO integrated intensity \citep{brunt04} between $\rm{V_{LSR}}$ $-25\,\rm km\,s^{-1}$ to $-15\,\rm km\,s^{-1}$ is shown in filled grey contours. The contour levels range from 10\% to 90\% of the peak value of $37.3\,\rm K\,km\,s^{-1}$. The Bolocam $1.1\,\mathrm{mm}$ \citep{ginsburg13} dust continuum intensity emission is overlaid as blue contours, ranging from 5\% to 95\% of the peak value of 2.0 Jy/beam. Sub-regions identified by \citet{kirsanova08} are shown by dashed boxes. Two diffuse regions, D1 and D2, used for magnetic field estimates of the cloud are shown as circles. The ellipses in the lower left corner represent the angular resolutions of the Bolocam ($33\arcsec$) and $^{13}$CO ($46\arcsec$) data.}
\label{fig9}
\end{figure}

Molecular clumps are cold, high-density structures often associated with star or star cluster formation \citep{evans99} and can be traced through their dust continuum emission. Determining clump physical properties and studying their magnetic field properties are needed to establish the dynamics of triggered star formation at smaller scales. We used Bolocam $1.1\,\mathrm{mm}$ \citep{ginsburg13} data to identify dense and massive clumps around the S235 H\,{\sc ii} region. Figure~\ref{fig9} shows the distribution of dust emission in blue contours and identifies the clumpy nature of the S235 complex. 

The Bolocam Galactic Plane survey \citep{aguirre11} used the customized algorithm Bolocat \citep{rosow10} to identify clumps and to generate catalogs of sources. The algorithm identified sources based on their signal compared to a local estimate of the noise. Regions with high signal are then subdivided into individual sources, based on the presence of local maxima within the region. These sources are expanded to include adjacent low-significance signal using a seeded watershed algorithm \citep{yoo04}. The properties of the sources are measured using the moments of the emission assigned to each of the identified sources.

\begin{figure}[!ht]
\epsscale{1.2}
\plotone{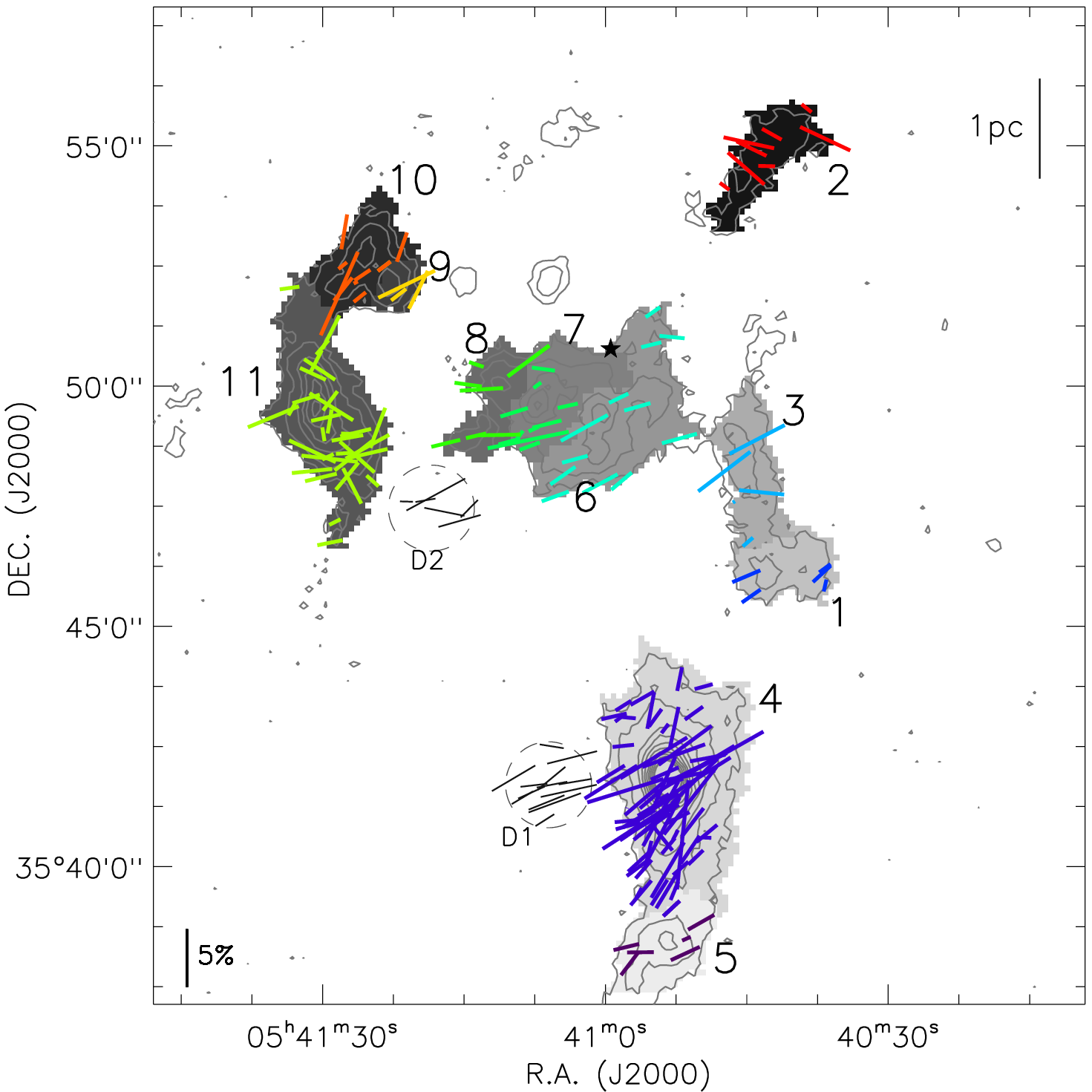} 
 \caption{The dense clumps identified in the Bolocam $1.1\,\mathrm{mm}$ dust emission are shown as zones of different grey shades. The contours and boundaries are same as in Figure~\ref{fig9}. The NIR embedded and background starlight polarization vectors are over-plotted in different colors to show the magnetic field orientations in each clump. The diffuse regions in the cloud, D1 and D2, are shown in dashed circles, also with their polarization vectors.}
   \label{fig10}
\end{figure}

We applied the Bolocat algorithm to the S235 data, using a detection threshold of 3$\sigma$ and an expanding parameter of 1.5$\sigma$, for an estimated $\sigma$ of $0.035\,\rm Jy/beam$. This produced 11 clumps having peak emission greater than 10$\sigma$. Figure~\ref{fig10} displays the identified clumps with their boundaries. Overplotted are embedded and background star polarization vectors representing magnetic field orientations. The 11 clumps span sufficient solid angles to encompass many starlight measurements each. The effective radii $R_{eff}$ of the clumps was estimated using a two-dimensional circular Gaussian fit to each clump. The Gaussian fit was then deconvolved by the Bolocam beam. The resulting values for $R_{eff}$ are listed in Table~\ref{tab4} and range from $0.26\,\mathrm{pc}$ to $0.75\,\mathrm{pc}$, with a mean radius of $0.5\,\mathrm{pc}$.

\subsubsection{Clump Masses and Column Densities}

The dust continuum emission offers a reliable tracer of the clump masses due to the low optical depths of the dust \citep{john06}. The total gas and dust mass, also called the isothermal mass $M_{iso}$, (due to the simplifying assumption that the dust can be characterized by a single temperature) within each molecular clump can be derived from the integrated $1.1\,\mathrm{mm}$ flux using the following equation \citep{hildebrand83}: 
\begin{equation}
M_{iso} \, = \, \frac{S_\nu \,D^2}{B_{(\nu,T_d)} \,\kappa_\nu} \;\; \mathrm{[g]}, \label{mass}
\end{equation} 
\noindent where $S_\nu$ is the integrated $1.1\,\mathrm{mm}$ flux in Jy, $D$ is the distance ($1.65\,\rm kpc$), $B_{(\nu,T_d)}$ is the Planck function at the dust temperature $T_d=20\,\mathrm{K}$ \citep{kirsanova14}, and $\kappa_\nu=0.0114\,\rm cm^{2}\,gm^{-1}$ is the dust absorption coefficient for a gas-to-dust ratio of 100 \citep{enoch06,bally10}.
Adopting these values, the equation simplifies to $M_{iso}=13.07 \times S_\nu \,D^2 \;\mathrm{[\Msun]}.$ 

The inferred clump masses range from 33 to 525 M$_{\odot}$ and are presented in Table~\ref{tab4}. The uncertainties were estimated by propagating the error values of integrated flux. The total gas mass of all 11 clumps is $\sim$1520 M$_{\odot}$. \citet{kirsanova14} estimated gas masses using NH$_{3}$ spectral lines, for a few sub-regions in the S235 complex. We find good agreement with their mass estimates, with a comparative mass ratio of $1.05\pm0.07$ between \citet{kirsanova14} values and this work. 

The column densities of the clumps are essential for assessing the effects of self-gravity and magnetic fields. We calculate two measures of the column density: the integrated column density $N(\mathrm H_2)^{int}_{1.1}$, and the average column density $N(\mathrm H_2)^{avg}_{1.1}$. The integrated column density is given by the following equation \citep{bally10}:
\begin{equation}
N(\mathrm H_2)^{int}_{1.1} = \frac{S_{\nu}}
{B_{(\nu,T_d)} \, \kappa_\nu \,\mu_{H_{2}} \,m_{\mathrm H} \,\Omega_\mathrm{beam}} \mathrm{[cm^{-2}]}
\end{equation}
\noindent where $S_{\nu}$, $B_{(\nu,T_d)}$, and $\kappa_\nu$ are defined in equation~\ref{mass}. $\mu_{H_{2}}$ is the mean molecular weight per hydrogen molecule considered to be 2.8 \citep{kauffmann08}, $m_{\mathrm H}$ is the mass of single hydrogen atom in grams, and $\Omega_\mathrm{beam}$ is the beam solid angle given by $\left(\frac{\sqrt{\pi}\,\,\rm FWHM} {206265 \times \sqrt{4\,ln\,2}}\right)^{2}$. $\Omega_\mathrm{beam}$ was calculated to be $2.9 \times 10^{-8}\,\rm sr$ for the Bolocam FWHM of 33$\arcsec$. Adopting these values and for a dust temperature $T_d=20\,\mathrm{K}$, the integrated column density equation reduces to $N(\mathrm H_2)^{int}_{1.1}=2.014 \times 10^{22}S_\nu\,\,\mathrm{[cm^{-2}]}$. 

The average column density for each clump can be calculated from its isothermal mass within a given aperture as \citep{dunham11}:
\begin{equation}
N(\mathrm H_2)^{avg}_{1.1} = \frac{M_{iso}}
{\mu_{H_{2}}\,m_{\mathrm H}\,\pi\,R_{eff}^2} \;\;\mathrm{[cm^{-2}]},
\end{equation}
\noindent where $M_{iso}$, $\mu_{H_{2}}$, $m_{\mathrm H}$ and $R_{eff}$ are defined earlier.

The estimated values of integrated and average column densities are listed in Table~\ref{tab4}. The quantity $N(\mathrm H_2)^{avg}_{1.1}$ is smaller than $N(\mathrm H_2)^{int}_{1.1}$ since it is averaged over the entire area of the clump. The $N(\mathrm H_2)^{avg}_{1.1}$ values range from $2.6\times10^{21}$ cm$^{-2}$ to $1.3\times10^{22}$ cm$^{-2}$, with a mean of $6.9\pm0.3\times10^{21}$ cm$^{-2}$ and a standard deviation of $2.5\pm0.2\times10^{21}$ cm$^{-2}$. These are consistent with the typical values for clumps in Galactic molecular clouds \citep{dunham11}.

\begin{table*}[t!]
\centering
\setlength{\tabcolsep}{0.11in}
\caption{Properties of Bolocam $1.1\,\mathrm{mm}$ dust clumps.}
\label{tab4}
\hspace*{-2cm}\begin{tabular}{ccccccccc}
\hline 
\hline 
Clump & R.A.   & DEC.  & $R_{eff}$ & $S_{\nu}$ & $M_{iso}$ & $M_{vir}$  & $N(\mathrm H_2)^{avg}_{1.1}$ & $N(\mathrm H_2)^{int}_{1.1}$ \\   
 ID  & (J2000) & (J2000) & (pc)  &(Jy) &  (M$_{\odot}$) &(M$_{\odot}$)  &(10$^{21}$ cm$^{-2}$) & (10$^{22}$ cm$^{-2}$)  \\   
\hline 	

1  & 85.17199 & 35.76990 & 0.40 & 1.49 (0.29) & 53 (12)  & 416 (71)   & 4.74 (1.13) & 3.01 (1.20) \\
2  & 85.17990 & 35.91203 & 0.48 & 1.25 (0.35) & 44 (13)  & 360 (126)  & 2.68 (0.83) & 2.52 (1.41) \\
3  & 85.18770 & 35.80451 & 0.42 & 2.02 (0.52) & 72 (20)  & 362 (103)  & 5.61 (1.62) & 4.08 (2.1)  \\
4  & 85.22087 & 35.69645 & 0.75 & 14.77 (0.49)& 525 (72) & 1080 (152) & 13.19 (1.81)& 29.75 (1.97) \\
5  & 85.22495 & 35.63886 & 0.43 & 2.45 (0.36) & 87 (17)  & 467 (102)  & 6.70 (1.34) & 4.93 (1.47) \\
6  & 85.24700 & 35.82111 & 0.68 & 5.25 (0.67) & 186 (34) & 1891 (497) & 5.59 (1.03) & 10.57 (2.7) \\
7  & 85.27268 & 35.82915 & 0.51 & 4.36 (0.43) & 155 (25) & 1204 (273) & 8.35 (1.38) & 8.78 (1.73) \\
8  & 85.29919 & 35.83027 & 0.35 & 1.40 (0.34) & 49 (14)  & 426 (110)  & 5.62 (1.57) & 2.82 (1.39) \\
9  & 85.34428 & 35.86830 & 0.26 & 0.94 (0.18) & 33 (8)   & 184 (40)   & 7.05 (1.63) & 1.91 (0.72)  \\
10 & 85.35845 & 35.87669 & 0.37 & 2.19 (0.30) & 77 (15)  & 336 (50)   & 7.93 (1.53) & 4.41 (1.23) \\
11 & 85.37528 & 35.82512 & 0.64 & 6.76 (0.42) & 240 (35) & 502 (57)   & 8.27 (1.21) & 13.63 (1.69) \\

\hline          
\end{tabular}
\begin{flushleft}
\tablecomments{Uncertainties are given in parentheses.}
\end{flushleft}
\end{table*}

Table~\ref{tab4} lists the derived physical properties of the 11 clumps. The table columns are ordered as follows: clump ID, equatorial coordinates R.A. and DEC obtained at the centroids of the clumps, clump effective radius, integrated flux, isothermal mass, virial mass (see Section~\ref{virial}), and the average column density.

\subsection{Molecular Cloud Distribution} \label{COcloud}

We used the optically thin $^{13}$CO (J = 1--0) line data \citep{brunt04} to obtain the molecular cloud properties. The analysis of the $^{13}$CO channel maps revealed that emission from the S235 complex was associated in the velocity range $-25\,\rm km\,s^{-1}$ to $-15\,\rm km\,s^{-1}$ \citep{kirsanova08,dewangan17}. The spatial distribution of the molecular gas was traced by integrating the intensity of the $^{13}$CO spectra over that velocity range. Figure~\ref{fig9} shows the distribution of $^{13}$CO integrated emission as grey scale contours. The molecular gas traces a shell-like envelope around the S235~Main region with a cavity towards the north. The gas is distributed inhomogeneously and shows several dense condensations. The peak emission for both the dust and the molecular data show close association (see Figure~\ref{fig9}), with the average dust to gas peaks offset by only $28\pm6\arcsec$. This indicates the identified dust clumps correspond to denser molecular gas.

Studies by \citet{kirsanova08,kirsanova14} compared emission from CS(2--1) (a biased tracer of dense gas) and $^{13}$CO(1--0) (a less biased tracer of all molecular gas) to show that the S235 molecular cloud is separated into different phases, depending on the component velocity. They identified three main phases: undisturbed quiescent gas ($-18\,\rm km\,s^{-1}< V_{LSR} < -15\,\rm km\,s^{-1}$); gas affected by the expanding H\,{\sc ii} region ($-21\,\rm km\,s^{-1}< V_{LSR} < -18\,\rm km\,s^{-1}$); and gas expelled from vicinities of the embedded young stars ($-25\,\rm km\,s^{-1}< V_{LSR} < -21\,\rm km\,s^{-1}$). Based on this, they suggested the H\,{\sc ii} region in the S235~Main is expanding largely towards the observer. They further divided the complex into different sub-regions, based on the locations of young star clusters, namely: East1; East2; Central; North-West; and Central-West. These sub-regions are highlighted by dashed boxes in Figure~\ref{fig9}. 

%Linewidths exceed thermal values, so non-thermal motions exist in the undisturbed quiescent gas,

In order to estimate the magnetic field properties in the diffuse part of the S235 cloud, we selected two circular regions (D1 and D2, see Figure~\ref{fig9}), that lacked any overlap with the dust clumps. The regions were chosen based on areas having high density of polarization stars covering both S235~Main and S235AB. The radius of each circular region was set equal to the mean effective radius of the 11 clumps. Since the diffuse regions lacked any prominent dust emission, we calculated their hydrogen column densities using $^{13}$CO data. We applied the column density equation given by \citet{simon01}:
\begin{equation}
N(\mathrm H_2)_{\rm CO} = 4.92\times10^{20} T_{mb}{\Delta}v \;\;\mathrm{[cm}^{-2}],
\end{equation}
\noindent where $T_{mb}$ is the peak antenna temperature in K and ${\Delta}v$ is the velocity FWHM line width in km~s$^{-1}$, both obtained from the $^{13}$CO spectra, corrected for main-beam efficiency of 0.48 \citep{jackson06}. The average column density $N(\mathrm H_2)^{avg}_{\rm CO}$ for the diffuse regions was calculated as the mean of the column densities for all pixels in each region. The $N(\mathrm H_2)^{avg}_{\rm CO}$ for D1 and D2, were $2.1\times10^{21}$ cm$^{-2}$ and $3.7\times10^{21}$ cm$^{-2}$, respectively.

\subsection{Magnetic Field Strengths} \label{Bfield}

The POS magnetic field strength can be estimated by the method proposed by \citet{davis51} and \citet{chandrasekhar53}, hereafter the DCF-method. This method relies on the assumption of equipartition between gas turbulent and mean magnetic energies. Hence, turbulent gas motions (dispersion in gas velocity) will lead to irregularities in magnetic field orientations which can be sensed via dispersions in polarization position angles. Following DCF, the POS magnetic field strength ($B_{pos}$) is estimated as 
\begin{equation}
B_{pos} = Q \sqrt{4\pi\rho} \,\frac{\sigma_{v}}{\delta \theta} \;\;[\mu\mathrm{G}],
\label{bpos_def}
\end{equation}
\noindent where $\rho$ is the gas mass volume density (in gm cm$^{-3}$), $\sigma_{v}$ is the gas velocity dispersion (in cm s$^{-1}$), $\delta \theta$ is the $PA$ dispersion (in radians) and $Q$ is a scaling factor. Several simulations \citep{heitsch01,ostriker01} have shown that $Q$ value may be approximated as 0.5, yielding reliable results for $B_{pos}$, provided the $PA$ dispersion is less than $25^{\circ}$. Adopting $Q=0.5$, Equation~\ref{bpos_def} reduces to
\begin{equation}
B_{pos} = 1.77\left(\frac{\sqrt{\rho}}{\rm{1\,gm\,cm^{-3}}}\right)\left(\frac{\sigma_{v}}{\rm{1\,cm\,s^{-1}}}\right)\left(\frac{\rm1\, rad}{\delta \theta}\right)\;\;[\mu\mathrm{G}],
\end{equation}

%We note that the POS magnetic field component consists of a large-scale ordered field and a turbulent part. Since the above DCF-equation depends only on the turbulent part, some sub-mm polarization studies \citep{pattle17,coude19} have accounted for this by using modified DCF-methods, namely structure functions \citep{hildebrand09} and auto-correlation functions \citep{houde09}. These methods calculate the angular dispersion function (polarization dispersion between every two vectors as a function of vector separation) to estimate the ratio of small-scale and uniform magnetic energies. These methods work best if the inferred polarization vectors are closely spaced and in large numbers. Our starlight polarizations do not provide sufficient stellar density to consistently apply these methods throughout the S235 field. Hence, we use the DCF-method as the main procedure for $B_{pos}$ estimates. 

\begin{figure}[!ht]
\epsscale{1.1}
\plotone{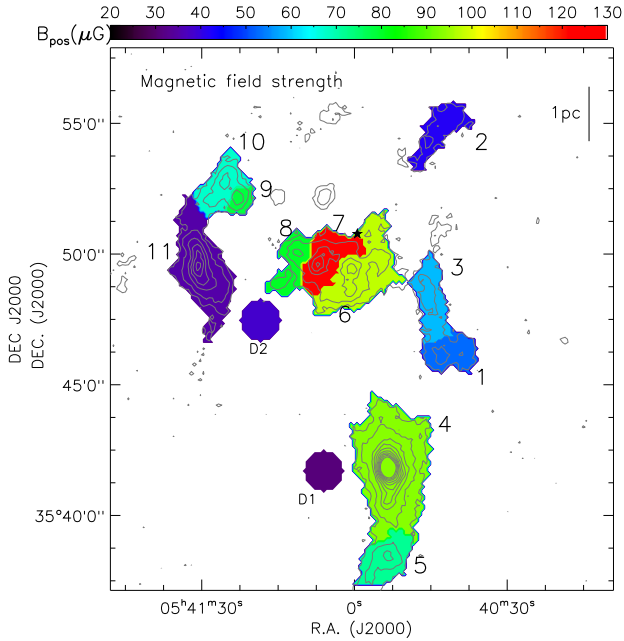} 
 \caption{Map of clump-averaged $B_\mathrm{pos}$ for the S235 complex, estimated using the DCF-method. The identified clumps and two diffuse regions of the cloud (D1 and D2) are highlighted in different colors corresponding to their average $B_\mathrm{pos}$ strengths. The color bar at the top indicates the range of $B_\mathrm{pos}$ values.}
   \label{fig11}
\end{figure}

We estimated $\delta \theta$ values from the background starlight polarization angles. Each dispersion value was calculated by taking the variance-weighted standard deviation of all the $PA$ values within each clump boundary. Since the $PA$ values have 180$^\circ$ ambiguity, de-aliasing was necessary \citep{hoq17, wang20}. This was carried out by stepping the aliasing window through increments of 1$^\circ$ between 0$^\circ$ to 180$^\circ$. The minimum standard deviation from all the de-aliased sets was chosen as the final dispersion value for each clump. 

The $\sigma_v$ values were obtained from the $^{13}$CO spectra associated with each clump. The spectrum at each pixel was fit by a Gaussian to obtain the standard deviation, i.e the dispersion value. The mean velocity dispersion for each clump was computed as the weighted average of all standard deviation values within the clump boundary. Although velocity dispersion is based on observed properties along the LOS, it is considered to be a valid value for the POS under the assumption that the turbulent motions are isotropic \citep[e.g.][]{hoq17}.

The mass volume density $\rho$, was determined from the average column density values. The volume number density was estimated as $n_{\rm H_{2}}=N(\mathrm H_2)^{avg}_{1.1} / \frac{4}{3} R_{eff}$ based on the assumption of a uniform density spherical clump. For the diffuse regions, $n_{\rm H_{2}}$ was estimated as $N(\mathrm H_2)^{avg}_{\rm CO} / ln $, where $ln$ is the cloud depth along the LOS. Depths of filamentary clouds can be characterized by assuming a cylindrical morphology with determination of their spine axis and inclination, but the S235 cloud is distributed inhomogenously around the shell of the H\,{\sc ii} region. Hence, we chose $ln$ to be an intermediate value greater than the clump sizes ($\sim1\,\rm{pc}$) but smaller than the diameter of the H\,{\sc ii} region shell ($\sim3\,\rm{pc}$). Based on this, an empirical mean of $ln=2\,\rm{pc}$ was taken for the diffuse regions. The mass volume density was finally calculated as $\rho = n_{\rm H_{2}}(\mu_{H_{2}})(m_{\mathrm H})$, where $\mu_{H_{2}}$ and $m_{\mathrm H}$ are defined earlier.

The magnetic field strength for the 11 clumps and the two diffuse regions was estimated using the DCF-method (see Table~\ref{tab5}). The uncertainties were estimated by propagating the uncertainties of the $PA$ dispersions, the velocity dispersions, and the volume densities. 

A map of clump-averaged $B_\mathrm{pos}$ is shown in Figure~\ref{fig11}, with a color bar indicating the mapping of color to magnetic field strength. The clumps have values ranging from 36 to $121\,\mathrm{\mu G}$, with a mean value of $\sim 65\,\mathrm{\mu G}$. The $B_\mathrm{pos}$ for clump 11 is less reliable since its $\delta \theta$ exceeds $25^{\circ}$. Clumps in the central region have relatively higher $B_\mathrm{pos}$ values due to their lower polarization position angle dispersions. The diffuse regions D1 and D2 have lesser values and average about $34\,\mu$G. Overall, these 13 values are similar to the typical values for magnetic field strengths in Galactic molecular clouds, which range from $1-100\,\mathrm{\mu G}$ for volume densities $10^{3}-10^{4}\,\rm cm^{-3}$, obtained from Zeeman observations \citep{bourke01,troland08} and polarized dust emission \citep{crutcher04,liu18}.

We note that the magnetic field strength estimates can be dominated by large errors despite use of most reliable observational parameters. These values are influenced by systematic uncertainties such as distance, inclination angle, flux calibration, dust temperature, and clump geometry \citep{crutcher04, liu18}. Some of these systematic uncertainties may be minimal in the S235 analyses. For example the distance has been precisely determined using \textit{Gaia} DR2 parallax of the ionizing star. The main contribution of systematic uncertainty is from the assumption of uniform density spherical clumps with a uniform dust temperature ($T_{d}=20$K). Therefore, the real uncertainties of the magnetic field strength estimates could be larger than the quoted values.
 
\subsection{Magnetic Energy vs Gravitational, Turbulent, and Thermal Energies}

\subsubsection{Mass-to-Flux Ratio} \label{subsec:mfrt}

The importance of the magnetic field with respect to gravity can be probed by considering the ratio of the mass of an object to its magnetic-critical mass $M_{\Phi,\rm crit}$. The magnetic-critical mass is defined by the condition that magnetic energy is equal to the gravitational energy for a region in magnetostatic equilibrium and is given as \citep{mouschovias1976,mckee07}:
\begin{equation}
M_{\Phi,\rm crit}= \frac{\Phi}{2\pi \sqrt{G}}, \label{magcrit}
\end{equation}
\noindent where $\Phi = B_{tot}\pi R_{eff}^2$, is the magnetic flux threading a region of a given radius.

\citet{crutcher04} presented these parameters into a dimensionless quantity called the normalized mass-to-flux ratio ($\overline{M/\Phi_{B}}$). This can be calculated from the magnetic field strength as:
\begin{equation}
 \overline{M/\Phi_{B}} = \frac{M_{iso}}{M_{\Phi,\rm crit}} = 7.6 \times 10^{-21} \frac{N(\mathrm H_2)}{B_{tot}},
\end{equation}
\noindent where $N(\mathrm H_2)$ is the average column density (in cm$^{-2}$) and $B_{tot}$ is the 3-dimensional total magnetic field strength (in $\mathrm{\mu G}$). Since linear polarization observations provide only $B_{pos}$, the LOS component needed to form the $B_{tot}$ is missing. \citet{crutcher04} showed that the total magnetic strength can be estimated as $B_{tot}=1.3 B_{pos}$, based on statistical analysis for an average field geometry and projection effects. We applied this for the $B_{tot}$ estimates.

Determining $\overline{M/\Phi_{B}}$ allows assessing the stability of a region. If $\overline{M/\Phi_{B}}>1$, the region is super-critical and gravitationally unstable, whereas when $\overline{M/\Phi_{B}}<1$, it is sub-critical and supported by magnetic fields. 

Based on the values of $N(\mathrm H_2)^{avg}$ and $B_{pos}$, we estimated the mass-to-flux ratios for the clumps and the diffuse regions (see Table~\ref{tab5}). The $\overline{M/\Phi_{B}}$ values range between 0.3 and 1.3 with a median of about 0.6. These values correspond to sub-critical states of the clumps, implying a dominant role of the magnetic field. The diffuse regions have values of 0.4 and 0.6, suggesting the magnetic field also dominates in the diffuse parts of the cloud.

\begin{table*}
 % \tiny
\centering
\setlength{\tabcolsep}{0.11in}
\caption{Magnetic field properties in the S235 complex.}
\label{tab5}
\hspace*{-2cm}\begin{tabular}{ccccccccc}
\hline 
\hline
 Clump & No. of & $\delta \theta$ & $\sigma _{v}$ & $n_{\rm H_{2}}$  &$B_{pos}$ & $\overline{M/\Phi_{B}}$& $\mathcal{M_A}$ & $\beta_{p}$ \\
 ID & stars &  (degree) & (km s$^{-1}$)  & (cm$^{-3}$) &($\mathrm{\mu G}$) &  & & ($\times10^{-2}$)\\
\hline
  1 &  5  &  21.3 (2.6) &  0.91 (0.06) &  2890 (690) &  50 (9)  &  0.5 (0.2) &  0.6 (0.1) &  4.5 (1.6) \\
  2 &  8  &  11.0 (1.7) &  0.55 (0.04) &  1350 (420) &  40 (9)  &  0.4 (0.2) &  0.3 (0.1) &  3.2 (1.4) \\
  3 &  5  &  15.4 (2.2) &  0.71 (0.08) &  3190 (920) &  57 (14) &  0.6 (0.2) &  0.4 (0.1) &  3.8 (1.8) \\  
  4 &  69 &  17.2 (1.7) &  0.99 (0.05) &  4270 (590) &  82 (11) &  0.9 (0.2) &  0.5 (0.1) &  2.5 (0.5) \\
  5 &  7  &  15.2 (2.3) &  0.81 (0.05) &  3790 (760) &  72 (13) &  0.5 (0.2) &  0.4 (0.1) &  2.9 (1.0) \\
  6 &  12 &  10.4 (1.4) &  1.03 (0.05) &  1980 (370) &  97 (16) &  0.3 (0.1) &  0.3 (0.1) &  0.8 (0.3) \\
  7 &  11 &  11.7 (1.5) &  1.02 (0.05) &  3960 (660) & 121 (19) &  0.4 (0.1) &  0.3 (0.1) &  1.1 (0.3) \\
  8 &  7  &  14.3 (2.1) &  0.82 (0.03) &  3860 (1080)&  78 (16) &  0.4 (0.2) &  0.4 (0.1) &  2.5 (1.0) \\
  9 &  4  &  15.8 (2.4) &  0.66 (0.02) &  6580 (1520)&  74 (14) &  0.6 (0.2) &  0.4 (0.1) &  4.7 (1.6) \\
 10 &  9  &  19.7 (2.7) &  0.82 (0.02) &  5170 (990) &  66 (11) &  0.7 (0.2) &  0.5 (0.1) &  4.8 (1.4) \\
11$^\dagger$ &  33 &  29.0 (5.7) &  0.85 (0.04) &  3130 (460) &  36 (8)  &  1.3 (0.4) &  0.8 (0.2) &  9.5 (3.2) \\
\hline 
D1 & 13 &  10.8 (1.5) &  0.77 (0.09) &  370 (110) &  30 (7)  &  0.4 (0.2) &  0.3 (0.1) &  1.6 (0.7) \\  
D2 & 6  &  17.1 (2.6) &  1.21 (0.09) &  630 (200) &  39 (9)  &  0.6 (0.3) &  0.5 (0.2) &  1.7 (0.8) \\
\hline          
\end{tabular}
\tablecomments{Uncertainties are given in parentheses.}
\tablenotetext{}{ $^\dagger$Clump 11 has $\delta \theta$ greater than $25^{\circ}$ and hence its magnetic field estimates are less reliable.}
\end{table*}

\subsubsection{Turbulent Alfv\'{e}nic Mach Number and Plasma Beta} \label{mach}

The relative importance of gas turbulence with respect to the magnetic field can be described by the turbulent Alfv\'{e}n Mach number $\mathcal{M_A}$, and is an important factor in determining the evolution of clumps.

The Alfv\'{e}nic Mach number can be estimated as: 
\begin{equation}
\mathcal{M_A} = \frac{\sigma_{\rm NT}}{\mathcal{V_A}},
\end{equation}
\noindent where the Alfv\'{e}n speed is given by $\mathcal{V_A}=B_{tot}/\sqrt{4\pi\rho}$ and $\sigma_{\rm NT}$ is the non-thermal gas velocity dispersion. The non-thermal velocity dispersion was estimated as $\sigma_{\rm NT}=\sqrt{\sigma_{v}^2-\sigma_{\rm T}^2}$, where $\sigma_{\rm T}$ is the thermal line broadening, taken to be $0.15\,\rm km\,s^{-1}$, using values obtained from NH$_{3}$ line observations \citep{kirsanova14,dewangan16}.

Models by \citet{padoan01} and \citet{naka08} suggest that sub-Alfv\'{e}nic ($\mathcal{M_A} \le$ 1) magnetic fields are strong enough to regulate gas motions and allow gas to move along the field lines to form high density structures. In the super-Alfv\'{e}nic case ($\mathcal{M_A} >$ 1), gas turbulence can efficiently change the magnetic field from ordered to a more random morphology.

The turbulent plasma beta parameter $\beta_{p}$ characterizes the importance of magnetic fields with respect to the thermal energy within a clump. It is defined as the ratio of thermal to magnetic pressure:
\begin{equation}
\beta_{p} = \frac{P_{therm}}{P_{mag}} = \frac{2c_{s}^2}{\mathcal{V_A}^2},
\end{equation}
\noindent where $c_{s}$ is the sound speed in the neutral medium, calculated as $c_{s}=\sqrt{\frac{k T_{g}}{\mu_{H_{2}} m_{H}}} = 0.24\, \rm km\,s^{-1}$, for an average gas temperature of $T_{g}=20\,\rm{K}$ \citep{kirsanova14}.

The estimated values of $\mathcal{M_A}$ and $\beta_{p}$ are listed in Table~\ref{tab5}. The clumps show values consistently below unity for both parameters indicating sub-Alfv\'{e}nic conditions, and implying a strong magnetic field with respect to both turbulence and thermal pressure.

\subsubsection{Virial Analysis} \label{virial}

A virial analysis was performed to check if the clumps are currently in gravitational equilibrium. The virial mass, considering thermal, turbulent, and magnetic energies, is given as
\citep{pillai11, liu18}:
\begin{equation}
M_{vir}=3\frac{R_{eff}}{G}\left(\frac{5-2i}{3-i}\right)\left(\sigma_{\rm NT}^2+c_{s}^2+\frac{\mathcal{V_A}^2}{6}\right)
\end{equation}
where $i$ is the spectral-index for the density profile $\rho(r)$ as a function of the distance ($r$) from the clump center, $\rho(r)=r^{-i}$. We adopt $i=1.8$ from the \citet{mueller02} analysis of 31 massive star forming clumps in other Galactic regions. The $\sigma_{\rm NT}, c_{s}$, and $\mathcal{V_A}$ terms are defined previously. The resulting virial masses for the clumps are listed in Table~\ref{tab4}. The clumps show consistently larger virial masses than their isothermal masses, indicating the regions are gravitationally stable against collapse. Therefore, the likelihood of formation of new star clusters in the S235 complex is currently low. 

%Alternatively, we can crosscheck the magnetic field strength when the region is in Virial equilibrium ($B_{\rm eq}$) with respect to observed with $B_{pos}$. The $B_{\rm eq}$ assuming equipartition of magnetic, gravitational, and kinetic energy is given as $B_\mathrm{EQ} = 2\sigma_{\rm NT}\sqrt{3\pi\rho}$ \citep{lada04}. The $B_{\rm eq}$ range from $18-30\,\mathrm{\mu G}$, with a mean value of $\sim20\,\mathrm{\mu G}$. These are lower by almost a factor of 3 compared to $B_{pos}$ and hence indicating the region to be magnetically dominated.

\section{Discussion} \label{sec5}

The analyses of 375 embedded and background starlight polarizations revealed the POS magnetic field orientations in the S235 complex. Estimates of clump magnetic field properties showed that the magnetic fields are dynamical important with respect to gravitational and turbulent energies. In this section, we discuss the properties of dust grain alignment, the distribution of magnetic field orientations around the H\,{\sc ii} region, and the role of magnetic fields in triggered star formation.

\subsection{Polarization Efficiency} \label{PE}

Properties of dust grain alignment with respect to the local magnetic field may be revealed through the polarization efficiency. It is defined as the ratio of the polarization percentage to the extinction (PE = P$_{\lambda}$/$A_{V}$). The grain alignment model of radiative torque alignment (RAT) \citep{laz07, hoang08} requires that anisotropic radiation imparts a net torque on the asymmetric dust grains. The dust grains then precess and align their rotational axes with the local magnetic field. A main prediction of RAT theory is that PE should decreases with increasing in $A_{V}$, since the radiation required to align the dust gets absorbed prior to reaching the denser regions. 

\begin{figure}[!ht]
\epsscale{1.2}
\plotone{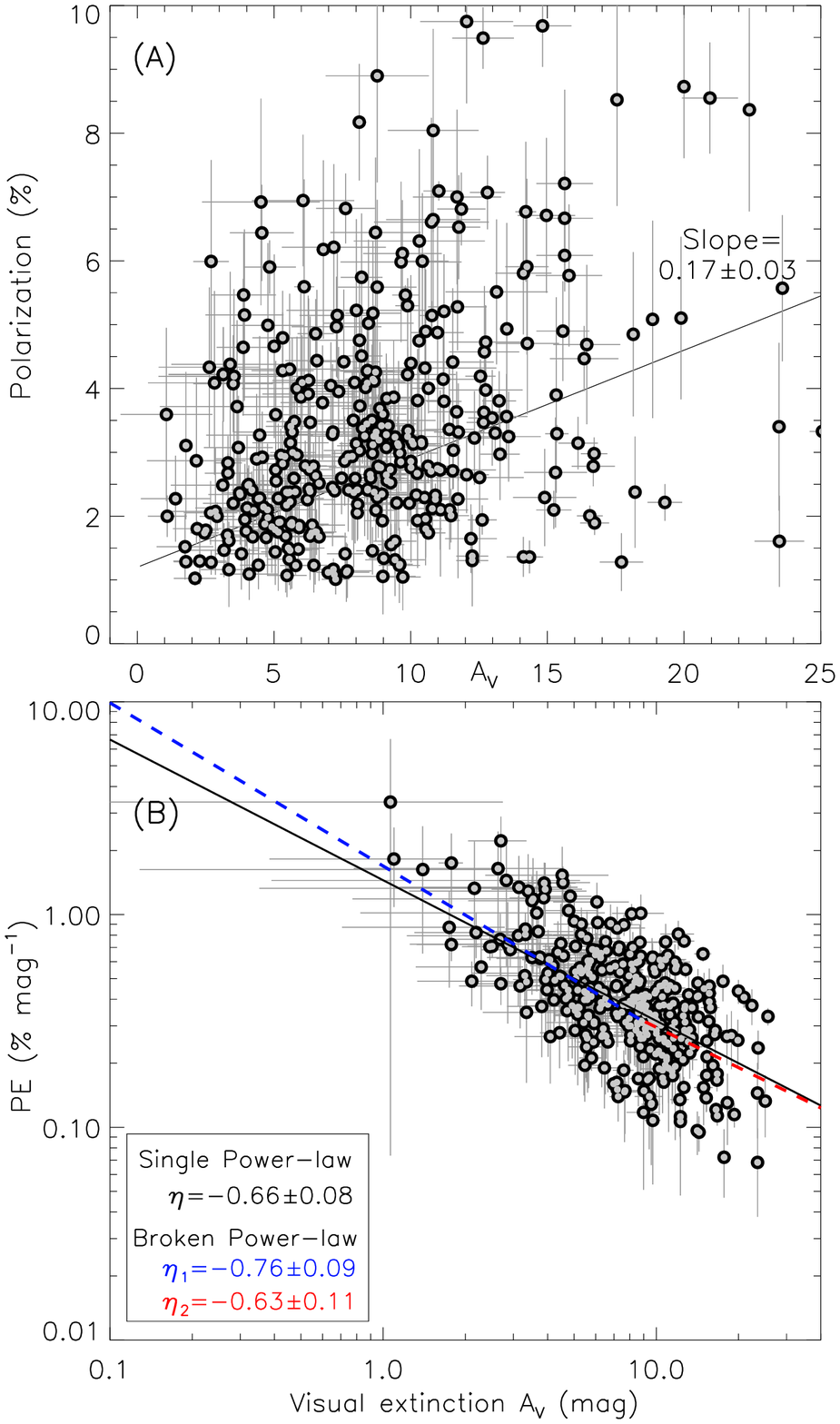} 
 \caption{\textit{Panel}(A): Plot of polarization percentage versus extinction for the 375 embedded and background stars in the S235 complex. A variance-weighted least-square fit gives a slope of $0.17\pm0.03$. \textit{Panel}(B): Log-log plot of polarization efficiency versus extinction for the same stars in \textit{Panel}(A). A single variance-weighted power-law fit on the values is shown in black solid line having an index of $\eta=-0.66\pm0.08$. A broken power-law fit is shown by the dashed blue and red lines, respectively. They have indices $\eta_1=-0.76\pm0.09$ and $\eta_2=-0.63\pm0.11$, respectively.}
   \label{fig12}
\end{figure}

We compared the $H$-band polarization percentage versus extinction for the embedded and background stars, and the results are shown in Figure~\ref{fig12}(A). The polarization values exhibit an increasing trend with extinction, especially for $A_{V}>3\,\rm mag$, reaching a maximum polarization of about $8-10\,\%$ for $A_{V}\sim15\,\rm mag$. The $P$ vs $A_{V}$ relation was fit with a variance-weighted least-square line, giving a slope of $0.17\pm0.03$. The positive slope is close to the value of 0.2 obtained by \citet{kusune15} for dense clouds at the peripheries of other H\,{\sc ii} regions. Other studies have shown a lower $P$ vs $A_{V}$ slope of about 0.1 \citep{arce98, hoq17} with predictions of decrease in polarization efficiency at large opacities in their clouds. The slope of $0.17\pm0.03$ found for the S235 complex may be attributed to typical dust grains ($0.1-0.5\,\mu$m) which remain efficiently aligned by radiation to substantial optical depths \citep[see review by][]{and15}. 

The relation between PE and $A_{V}$ has been characterized for other clouds using a power-law behavior (${\rm PE}\propto A_{V}^{\eta}$) mostly to match predictions of RAT theory. Several studies have shown the power-law indices range from $\eta=-0.3$ to $-0.7$ \citep{whittet08,chapman11,alves14,cashman14} for medium to high opacity clouds. Figure~\ref{fig12}(B) compares the PE vs $A_{V}$ for embedded and background stars in the S235 complex. We fit a single variance-weighted power-law to the values and obtained a slope of $\eta=-0.66\pm0.08$. The PE tends to be greater for stars with low extinctions, up to $A_{V}\sim5\,\rm mag$, and becomes lesser for higher extinctions. This might indicate that the high PE values are associated with grains lying in the envelopes of the clumps.

\citet{whittet08} used a large sample of $K$-band polarimetry towards Taurus and Ophiuchus clouds to find a single power-law index of $-0.52\pm0.07$ over the full range of $A_{V}=0-30\,\rm mag$. However, their predictions of RAT modeling suggested a break point should exist at higher extinctions. \citet{alves14} showed that the PE in Pipe nebula exhibits a break point at $A_{V}\sim10\,\rm mag$, interpreted as due to a mixed distribution of grain sizes from cloud to dense cores. \citet{wang17} performed a similar study with IC5146 cloud and found the PE in $H$-band drops steeply ($\eta=-0.95\pm0.30$) up to $A_{V}\sim3\,\rm mag$, but then changes to a more softer value ($\eta=-0.25\pm0.10$) for greater $A_{V}$. 

To test whether the PE vs $A_{V}$ relation has a break point for the S235 complex, we fit a broken power-law of the form: 
\begin{equation}
PE=
\begin{cases}
\alpha_1 {A_V}^{\eta_1} & : A_{V} \leq x_{break}\\
 \alpha_2 {A_V}^{\eta_2} & : A_{V} > x_{break},\\
\end{cases}
\end{equation}
where $\alpha_2$=$\alpha_1 x_{break}^{\eta_1-\eta_2}$. The parameters, $\alpha_1$, $\eta_1$, $\eta_2$, and $x_{break}$, were taken as free parameters for the broken power-law fit.

We find the slopes of broken power-law to be $\eta_1=-0.76\pm0.09$ and $\eta_2=-0.63\pm0.11$. These are represented by dashed blue and red lines in Figure~\ref{fig12}(B). The difference in the slopes for the broken power-law case and the single power-law case is small. This indicates that the PE vs $A_{V}$ behavior likely changes systematically over the entire $A_{V}$ range in the S235 complex. This general decrease in PE within the cloud may arise if the grain size distribution are same in the diffuse and denser regions. This requires no significant grain growth, which if present would be expected to display a flatter index at large $A_{V}$. 

%The grain alignment in the S235 complex is also expected to be affected by the radiation from nearby young massive star(s). \citet{cashman14} have shown that PE decreases systematically with projected distance from the radiating source, with their PE vs $A_{V}$ index $\eta=-0.74\pm0.07$. The clumps in the S235 complex are located in projection to be less than $3\,\rm pc$ from the central ionizing O9.5V star. Our index of $0.69\pm0.09$ is similar to \citet{cashman14} and could indicate radiation from massive stars have a role in grain alignment. 

%The H\,{\sc ii} region in S235 is expanding largely towards the LOS of the observer \citep{kirsanova08,kirsanova14} and clumps are located at the periphery. Hence, grain alignment from radiation of the central star should be significant at the far side of the clumps. This implies that the high opacity regions of the clumps can have well aligned dust grains for our case.

A possible caveat in the $\eta$ estimates arises from the uncertainty in the $A_{V}$ calculations. This is due to the estimation of $A_{V}$ by assumption of intrinsic stellar colors without knowing their spectral types. However, this should not significantly change the PE vs $A_{V}$ trend. Hence, we assume the background star polarization is tracing the magnetic field within the clumps as predicted by RAT theory.

\subsection{Magnetic Field Distribution}

\subsubsection{Large-scale Magnetic Field}

The global polarization orientations shown in Figure~\ref{fig8} reveal an ordered magnetic field throughout the S235 cloud. Studies have suggested that molecular clouds tend to preserve large-scale magnetic field orientations \citep[see][]{lihb14,clemens18} for which gravitational flow of gas is unconstrained along the magnetic field lines. We used the \textit{Planck} polarized dust emission, rotated 90$^{\circ}$ to reveal the large-scale field structure. Figure~\ref{fig13} shows a magnetic field map using NIR and \textit{Planck} data. Both data sets exhibit similar large-scale patterns, with their magnetic field orientations being mostly parallel to the Galactic plane. Several polarization surveys show similar distributions, where polarization angles tend to present an overall plane-parallel orientation \citep{mat70,heiles05,clemens12a}. The S235 complex is situated in the Milky Way Perseus spiral arm and extends to a latitude of $b\approx+2.8^{\circ}$. At these latitudes, the bulk of gas still remains part of the spiral arms and the large-scale magnetic fields are believed to have their origin due to differential rotation in the Galactic disk \citep{beck96,beck15}. 

We quantitatively compared the magnetic field orientations from the NIR and \textit{Planck} data sets. The mean polarization angle in the NIR data is $114\pm4^{\circ}$ and for the \textit{Planck} data (rotated 90$^{\circ}$) is $121\pm2^{\circ}$. The difference between both the data sets is small, indicating that polarization due to dust absorption and dust emission are tracing the same component of magnetic field at larger scales. At smaller scales, the NIR and \textit{Planck} data differ. We compared the polarization angles for clumps 4 and 11, which have sizes similar to the \textit{Planck} effective beam size of $5\arcmin$ ($\sim2\rm pc$). Other clumps were excluded as their sizes are smaller than the \textit{Planck} beam. Clump 4 shows a strikingly similar value between NIR ($123\pm3^{\circ}$) and \textit{Planck} ($122\pm1^{\circ}$). Whereas, clump 11 has a significant difference, with values $110\pm4^{\circ}$ and $148\pm2^{\circ}$, respectively. The discrepancy in results can be due to the several factors of which turbulence from stellar feedback can significantly alter the small-scale magnetic field orientations. Hence, \textit{Planck} polarization is useful for mainly sampling the large-scale ordered magnetic fields.

\subsubsection{Magnetic Field in the Shell}

\begin{figure}[!t]
\epsscale{1.2}
\plotone{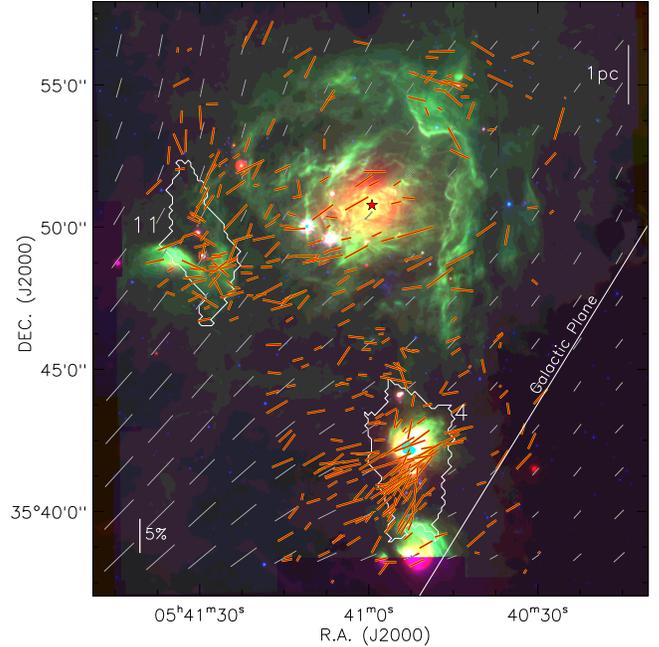} 
 \caption{Large-scale magnetic field orientations in the S235 complex. Background image is three-color composite map using \textit{Spitzer} MIPS and IRAC bands: 24$\mu$m (red), 8$\mu$m (green) and 3.6$\mu$m (blue). The \textit{Planck} polarizations, rotated $90^{\circ}$, tracing the large-scale magnetic field are shown as grey vectors. The NIR polarization are shown in red-yellow vectors. The direction of the Galactic plane is shown as a solid white line. Clumps 4 and 11 used in comparing NIR and \textit{Planck} polarizations are highlighted by their boundaries.}
   \label{fig13}
\end{figure}

The most prominent result seen in the $H$-band polarization observations is the curved magnetic field morphology around the S235~Main region (see Figures~\ref{fig8} and~\ref{fig13}). This pattern coincides with the location and morphology of the mid-infrared PDR shell that encompasses the expanding H\,{\sc ii} region. The shell is created when the expanding H\,{\sc ii} region pushes the surrounding material via radiation and thermal pressure. As the ionization front expands, it causes inhomogeneities and changes in density of the material. Hence, the gas around the shell gets compressed and is seen tracing a shell-like morphology in Figures~\ref{fig9}. Since the magnetic field is coupled to the gas, the field lines are dragged in this process, causing them to follow the spherical pattern of the shell.

Several numerical simulations incorporating magnetic fields \citep{krumholz07,arthur11} have shown curvature in magnetic field orientations due to expansions of H\,{\sc ii} regions. This has also been revealed in observational studies \citep[e.g.,][]{santos14,chen17,dew18}, where magnetic fields are found to be aligned tangentially to H\,{\sc ii} region shells. The curvature of field lines in S235~Main are mainly at the edges of the shell and correspond to distance of about $2.4\,\rm pc$ from the ionizing star. A typical bubble/shell like H\,{\sc ii} region consists of an inner ionized region surrounded by a PDR envelope with a thickness of about 0.2 to $0.4\,\rm pc$ \citep{church06}. Hence, the magnetic field distribution is mainly affected at the outer layers, where the ionization front interacts with the ambient molecular gas. 

Recent studies on the structure of the S235~Main H\,{\sc ii} region using ionized and molecular emission suggested that the molecular material in the line-of-sight is mainly distributed in the rear and side walls of the PDR shell \citep[see Figure 8 of][]{anderson19, kirsanova20}. This indicates that the magnetic field directions traced in clumps 6, 7 and 8 are associated to the regions behind the ionizing star.

\citet{arthur11} showed with simulations that the H\,{\sc ii} region expands largely parallel to the ambient magnetic field and over-dense regions are created at the edges of the elongated end caps. The magnetic field lines in S235~Main show similar distribution with slight elongation parallel to the large scale Galactic magnetic field. Additionally, the north-west and south-east regions are associated to dense molecular material (clumps 2 and 11), which appear to be projected at the outer edge of the H\,{\sc ii} region. This observational feature correlates well with the simulations and supports the prediction that the ionization front compresses the surrounding neutral gas into denser regions. These dense regions have been observed to harbor number of YSOs \citep{kirsanova14, dewangan16}. Hence, the formation of these YSOs are attributed to the effect of the expanding H\,{\sc ii} region.

\subsection{The Role of Magnetic Fields in Triggered Star Formation}

Triggered star formation has normally been considered in the context of weak magnetic field models \citep{crutcher12}, where supersonic turbulent flows dominate the evolution of molecular clouds. On the other hand, strong magnetic field models \citep{mouschovias1991,mouschovias1999} suggest molecular clouds are supported by well ordered magnetic fields through their outward pressure, which later quasi-statically dissipate via ambipolar diffusion \citep{mestel56}. Eventually on core scales, self-gravity overcomes magnetic forces, inducing collapse and the formation of stars \citep{mouschovias2006}. 

The S235 complex has been studied extensively to seek evidence for triggered star formation \citep{dewangan11,kirsanova14,dewangan16,dewangan17}. These studies have found large numbers of YSO clusters in the molecular gas. The studies have comprehensively arrived to the conclusion that YSO clusters are the result of positive stellar feedback driven by the expansion of the S235~Main H\,{\sc ii} region. The S235AB and S235C regions have also been observed to contain number of YSOs. \citet{felli06} proposed that these clusters could have formed due to the expanding H\,{\sc ii} region from massive star in S235A. However, other studies indicate that the region associated to S235A is a young ultra-compact H\,{\sc ii} region of about $0.2\,\rm Myr$ \citep{dewangan11} and lacks energetics to trigger new stars in the surrounding molecular core \citep{kirsanova14}. Hence, we focus our discussion of magnetic fields and triggered star formation mainly towards the S235~Main region.

\begin{figure}[!ht]
\epsscale{1.1}
\plotone{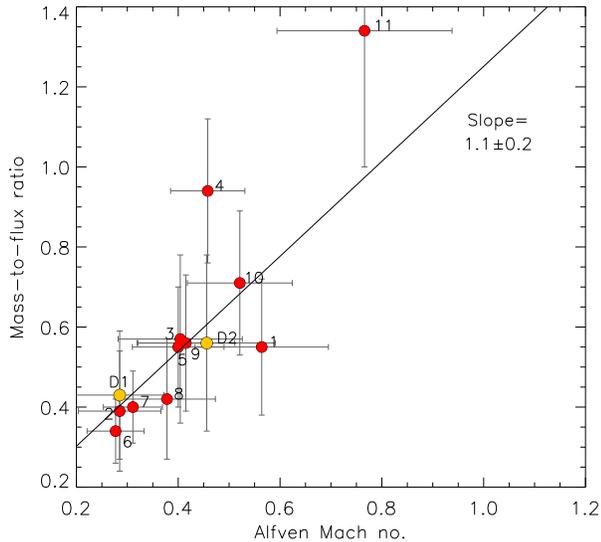} 
 \caption{Plot of normalized mass-to-flux ratio versus Alfv\'en Mach number for the clumps (red points) and diffuse regions (yellow points) of the S235 complex. The clump numbers are highlighted next to each data point. There is a tight linear dependence of both quantities, except for clump 11 having less reliable values due to its large polarization position angle dispersion (see Section~\ref{Bfield}). A variance-weighted least-square fit gives a slope of $1.1\pm0.2$.}
   \label{fig14}
\end{figure}

Several observational studies find molecular clouds to be magnetically super-critical \citep{crutcher99, crutcher12}. In contrast, the S235 complex is magnetically sub-critical. The three parameters for criticality: $\overline{M/\Phi_{B}}$, $\mathcal{M_A}$, $\beta$, estimated from our observations, show values less than unity. The individual clumps show sub-critical states and are not in virial equilibrium. The Alfv\'en Mach number values suggest the clumps are sub-Alfv\'{e}nic and hence the role for turbulence is minimal. These results indicate strongly magnetized conditions prevail, both in the dense clumps and diffuse regions of the S235 complex. We compared the values of the mass-to-flux ratio to the Alfv\'en Mach number for the clumps and diffuse regions, as shown in Figure~\ref{fig14}. There is a tight linear correlation between both quantities, where $\overline{M/\Phi_{B}}$ increases for increasing $\mathcal{M_A}$. A variance-weighted least-square fit returns a slope of $1.1\pm0.2$. The positive linear slope indicates the magnetic fields dominate equally against both gravity and turbulence.

In order to understand how triggered star formation proceeds in the S235~Main in the presence of strong magnetic fields, we consider the evolution, fragmentation, and diffusion time-scales of the complex. Before the expansion of the S235~Main H\,{\sc ii} region, the strong magnetic fields in the parental cloud likely helped keep the region in gravitational equilibrium. As the H\,{\sc ii} region expanded, it created overdense regions, since supersonic turbulence from the shock the fronts compressed the molecular gas. The magnetic fields at this stage did not provide sufficient pressure to counter the ionization fronts. This period lasted short for about $0.1\,\rm Myr$ \citep{elmegreen77}, after which it led to core collapse, forming low-mass stars or even high-mass stars, as predicted by competitive accretion \citep{tan14}. The turbulence driven by the shock fronts decayed rapidly with the regions transitioning into sub-Alfv\'{e}nic states. The magnetic fields in the envelopes still preserved their sub-critical natures. This describes the current conditions present in the S235~Main region.

\citet{bisbas11} carried out numerical simulations of triggered star formation, where pre-existing dense condensations were compressed by the pressure of the ionized gas. Their results showed that star formation mainly occurs during the initial phases of the interactions of shock fronts with quiescent ambient molecular gas, when the ionizing photon fluxes are high (of the order $10^{9}-10^{11}\,\rm cm^{-2}\,s^{-1}$). They also demonstrated that the sequence of triggered star formation occurs on a rapid time-scale, of about $0.19$ Myr. \citet{mackey11} showed that the energy from photo-ionization at the boundaries of an H\,{\sc ii} region is significantly stronger than the magnetic fields present there. Hence, turbulence is the dominant driver at the initial stages that leads to gas compression in the forms of pillars and globules. \citet{arthur11} studied the evolution of an H\,{\sc ii} region in a turbulent, magnetized molecular cloud and found that after the expansion of the H\,{\sc ii} region the magnetic fields in the shell are amplified and slow the formation of new stars.

These modeling results, considered in the context of our new observations can be interpreted as follows. The first generation of triggered stars in the S235~Main formed rapidly from the effects of expansion of the H\,{\sc ii} region. The formation of subsequent generations of stars from the remaining dense gas and envelope has been slowed by the strong magnetic fields present, as revealed in our observational results. Hence, turbulence from stellar feedback plays positive roles during the initial stages of triggered star formation, followed by magnetic fields regulating the star formation process thereafter. 

\begin{figure*}[!ht]
\epsscale{1.15}
\plotone{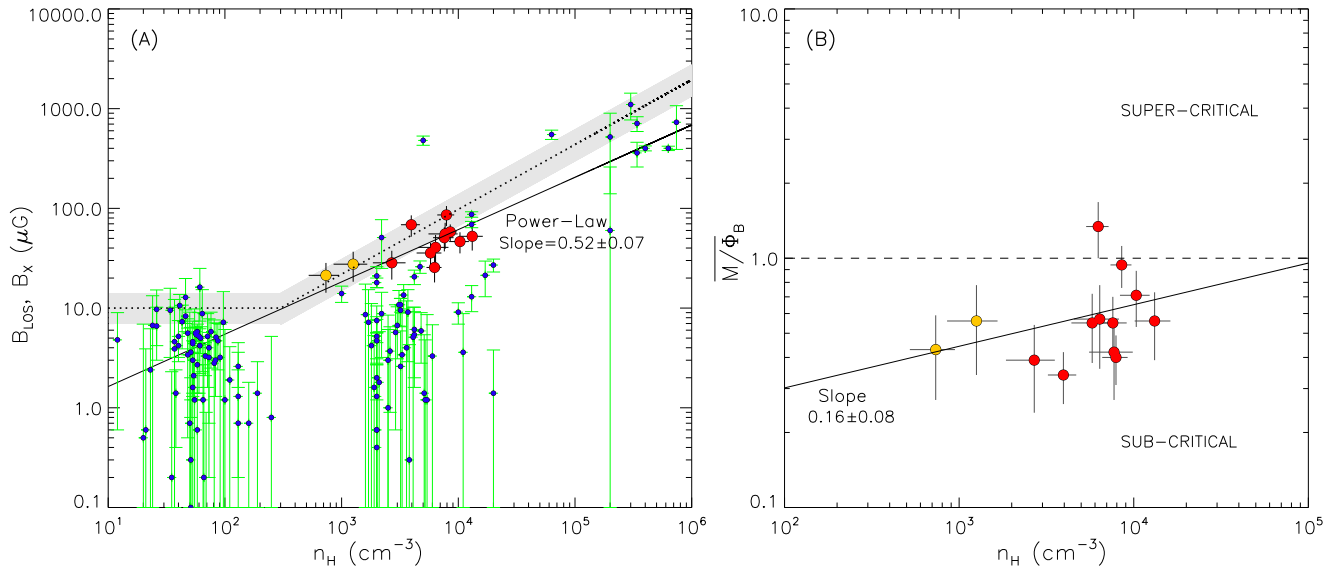} 
 \caption{\textit{Panel}(A): Log-log plot of one-dimensional magnetic field strength ($B_{\rm X}$) against atomic hydrogen volume density ($n_{\rm H} = 2\,n_{\rm H_{2}}$) for clumps (red points), and diffuse regions (yellow points) of the S235 complex. $B_{\rm X}$ is obtained by scaling $B_{pos}$ by $1/\sqrt{2}$. The $B_{\rm LOS}$ values and limits from \citet{crutcher10} are shown as blue points with green error bars. The dotted black line with grey shading (index of $0.65$) shows the maximum $B_{\rm LOS}$ with $1\sigma$ uncertainties for the \citet{crutcher10} Bayesian analysis. The solid black line shows the single power-law fit to the S235 values, which gives an index of $\kappa=0.52\pm0.07$. \textit{Panel}(B): Log-log plot of normalized mass-to-flux ratio ($\overline{M/\Phi_{B}}$) versus atomic hydrogen volume density for the S235 clumps (red points), and diffuse regions (yellow points). The black dashed line represents unity in $\overline{M/\Phi_{B}}$, differentiating sub-critical and super-critical states. A single power-law fit to the values gives a slope of $0.16\pm0.08$ and is shown by solid black line.}
   \label{fig15}
\end{figure*}

\subsubsection{Magnetic Field vs Density} \label{subsec:nhvsb}

The relation between magnetic field strength and gas density ($B_{tot}\propto n_{\rm H}^{\kappa}$) can be used to predict the role of magnetic fields in the formation of new clumps. If magnetic fields are important, dense regions must form primarily from flows of weakly ionized, but mostly neutral gas along the field lines. \citet{fiedler93} carried out detailed numerical simulation of strong field models with ambipolar diffusion driven cloud contraction and obtained $\kappa\approx0.47$. \citet{crutcher99} used Zeeman observations of 27 clouds and obtained $\kappa=0.47$ for a wide range of densities from $10^{3}-10^{6} \,\rm cm^{-3}$, thus favoring the predictions of ambipolar diffusion driven star formation. Conversely, \citet{crutcher10} employed Bayesian statistical analysis of observational data for a wide variety of individual cloud observations, including upper limits, and found $\kappa=0.65$ for gas number densities greater $300\,\rm cm^{-3}$. They suggested that a larger index is an indication of isotropic core contraction that is not likely regulated by magnetic fields. Several observational studies of resolved magnetic field maps of individual clouds have found higher indices of $\sim0.7$ \citep{march12,hoq17}. 

Here, the S235 clumps show a power-law scaling in the relation between mean one-dimensional magnetic field strength ($B_{\rm X}=B_{pos}/\sqrt{2}$) and atomic gas number density ($n_{\rm H}$), as shown in Figure~\ref{fig15}(A). A single power-law fit gives a slope of $\kappa=0.52\pm0.07$. This slope is shallower than the 0.65 index of \citet{crutcher10} and implies that the S235 clumps are likely influenced by magnetic fields. \citet{lihb15} studied magnetic field properties in NGC~6334 from scales of 0.01 to 10 parsecs. They obtained a slope of $\kappa=0.41$ and suggested molecular fragmentation is anisotropic, with magnetic fields playing crucial role in that cloud fragmentation. \citet{wang20} carried out a detailed study of the IC5146 cloud using a Bayesian approach and found a slope of $\kappa=0.5$. They also found the probability density function for having a strong magnetic field strength was narrow and Gaussian. The S235 slope correlates well with the aforementioned studies and demonstrates the S235 clumps are likely regulated by magnetic fields through ambipolar diffusion. 

One of the important predictions of the ambipolar diffusion mechanism is the increase in the mass-to-flux ratio from the sub-critical to the super-critical state with increasing density \citep{mouschovias1991, mouschovias1999}. This occurs since the drift of neutral gas into the core increases the density of the region without an equally significant increase in the magnetic flux. We compared $\overline{M/\Phi_{B}}$ against $n_{\rm H}$ for the S235 regions as shown in Figure~\ref{fig15}(B). There is a positive correlation between both quantities with clumps that are more dense tending to have higher $\overline{M/\Phi_{B}}$. A single power-law fit gives a slope of $0.16\pm0.08$. Extrapolating the slope (not shown in the figure), we infer a change of $\overline{M/\Phi_{B}}$ from sub-critical to super-critical for densities greater than $10^5\,\rm cm^{-3}$. At these large densities, most of the gas exists as dense cores that are gravitationally unstable and in turn have super-critical states. A similar study was carried out by \citet{hoq17} for the G28.23 dark cloud who found the critical density to be around $300-600\,\rm cm^{-3}$, which is comparable to densities found in semi-diffuse molecular regions. \citet{heitsch14} showed that the transition from sub-critical to super-critical is more efficient at high densities and on dense core size scales. 

The main concern with star formation solely regulated by ambipolar diffusion derives from the comparison of triggered star formation time-scales with ambipolar diffusion time-scales. The latter process gives rise to slow star formation \citep{shu87} which appear inconsistent with the typical ages of YSOs in molecular clouds \citep{hartmann01}. Instead, accretion, along magnetic field lines, but via shocks, appears to be a likely mechanism to tip cores over from being sub-critical to super-critical \citep{heitsch14}. Moreover, turbulent magnetic reconnection \citep{lazarian99,lazarian12}, although suggested as a means of speeding the star formation process, also appears too slow. Hence, the triggered star formation in S235 is more likely the combination of shock compression, turbulent motions, and ambipolar diffusion.

\section{Conclusions} \label{sec6}

We have studied the magnetic field properties in the S235 complex using new near-infrared $H$-band starlight polarimetric observations in combination with archival $1.1\,\mathrm{mm}$ dust continuum emission, $^{13}$CO molecular spectral line data and \textit{Planck} polarized dust emission. The main results of the study are:

\indent 1. The plane-of-sky magnetic field orientations were traced using 375 bona fide embedded and background starlight polarization measurements. The contributing stars were selected through the evaluation of \textit{Gaia} DR2 distances and extinctions derived from 2MASS and UKIDSS colors. The magnetic field orientations showed a curved morphology outlining the (spherical) shell created by the central star's ionized emission. This indicates the magnetic field coupled with the gas, is pushed by the expansion of the H\,{\sc ii} region. 

\indent 2. The large-scale magnetic field orientation inferred from \textit{Planck} data was found to be mostly parallel to the Galactic plane.

\indent 3. We identified 11 dense clumps in the S235 complex using $1.1\,\mathrm{mm}$ dust emission. The clump masses were estimated to be $33-525\,\rm M_\odot$, with clump averaged H$_2$ column densities of about $6.9\times10^{21}$ cm$^{-2}$. The $^{13}$CO spectral line data traced the S235 cloud in the velocity range $-25\,\rm km\,s^{-1}$ to $-15\,\rm km\,s^{-1}$, with the projected distributions of molecular gas and dust emission showing clear associations.

\indent 4. We estimated the POS magnetic field strength for the clumps using the Davis and Chandrashekar-Fermi method. The values ranged between $36-121\,\mathrm{\mu G}$. The estimated mass-to-flux ratios show the clumps to be magnetically sub-critical ($\overline{M/\Phi_{B}}\sim0.6$). The turbulent Alfv\'{e}n Mach number and plasma beta were also found to be less than unity ($\mathcal{M_A}\sim0.5$, $\beta_{p}\sim0.04$), indicating a strongly magnetized region.

\indent 5. The dust grain alignment with respect to the local magnetic field was studied through analysis of polarization efficiency. The observations reveal a negative power-law dependence for polarization efficiency versus extinction, with an index of $-0.66\pm0.08$. These results favor the RAT theory and indicate background starlight polarization traces the magnetic field within the clumps.

\indent 6. Comparison of magnetic field strength with density shows a power-law dependence, with an index of $0.52\pm0.07$. This is consistent with values obtained for strong field models driven by ambipolar diffusion cloud contraction. The dependence of mass-to-flux ratio on density shows a positive correlation, with a $\overline{M/\Phi_{B}}$ change from sub-critical to super-critical for densities greater than $10^5\, \rm cm^{-3}$. These results further confirm the predictions of strongly magnetized cloud evolution.

In summary, the S235 complex is revealed to be a strongly magnetized region in which the expansion of an H\,{\sc ii} region influences the magnetic field orientations. The star formation history indicates that stellar feedback may have initially played a positive role in triggering new stars. However, as the H\,{\sc ii} region evolved, the magnetic fields become dynamically important enough to regulate later global star formation.


\acknowledgements{The authors like to thank the anonymous referee for a detailed and
thoughtful review that has improved the scientific contents of the paper. This research has been supported by the European Research Council advanced grant H2020-ER-2016-ADG-743029 under the European Union’s Horizon 2020 Research and Innovation program. R.D acknowledges INAOE and CONACyT-Mexico for SNI grant (CVU 555629). A.L. acknowledges financial support by CONACyT-Mexico through project CB-A1S54450. JM acknowledges funding from a Royal Society-SFI University Research Fellowship (14/RS-URF/3219). We thank all the OAGH staff for their help in conducting observations with POLICAN. This research was conducted in part using the Mimir instrument, jointly developed at Boston University and Lowell Observatory and supported by NASA, NSF, and the W. M. Keck Foundation. The analysis software for Mimir data were developed under NSF grants AST 06-07500, 09-07790, 14-12269, and 18-14531 to Boston University. This work makes use of data products from the Two Micron All Sky Survey, which is a joint project of the University of Massachusetts and the Infrared Processing and Analysis Center/California Institute of Technology, funded by NASA and NSF. The authors acknowledge the use of SAOImage DS9 software which is developed with the funding from the Chandra X-ray Science Center, the High Energy Astrophysics Science Archive Center and JWST Mission office at Space Telescope Science Institute.}

\end{document}